%% file: WorkshopReport.tex
\documentclass[12pt,tightenlines,superscriptaddress]{revtex4PATCH}
\usepackage{latexsym}
\usepackage{graphicx}
\usepackage{url}
\usepackage{amssymb,amsmath,amsfonts,amstext,graphics, multirow}
\usepackage{graphicx}
\usepackage{epstopdf}
\usepackage{url}
\usepackage{appendix}
\usepackage{color,slashed}
\newcommand{\gsim}{\lower.7ex\hbox{$\;\stackrel{\textstyle>}{\sim}\;$}}
\newcommand{\lsim}{\lower.7ex\hbox{$\;\stackrel{\textstyle<}{\sim}\;$}}

\def\LL{{\cal L}}

\def\Ed{(Editor)} 
\newcommand{\comments}[1]{} 
\newcommand{\authorcite}[2]{[{\tt #1}: #2]}

\newcommand{\hc}{\text{ h.c. }}

\newcommand{\MET}{\mbox{$E_T\hspace{-0.237in}\not\hspace{0.18in}$}}

\newcommand{\bef}{\begin{figure}[htbp]\begin{center}}
\newcommand{\eef}{\end{center}\end{figure}}

\newcommand{\bea}{\begin{eqnarray}}
\newcommand{\eea}{\end{eqnarray}}

\begin{document}

\pagestyle{plain}
\title{\Large Simplified Models for LHC New Physics Searches}

\input{authorlist.tex}

\input{abstract.tex}

\maketitle
\newpage

\tableofcontents

\newpage

\input{introduction.tex}

\newpage

\input{longexamples.tex}

\newpage

\input{jets.tex}

\newpage

\input{heavy.tex}

\newpage

\input{leptons.tex}

\newpage

\input{photons.tex}

\newpage

\input{exotica.tex}

\newpage

\input{conclusions}


\input{acknowledgments.tex}


\bibliography{WorkshopReport}

\end{document}

%% file: authorlist.tex
\author{Daniele Alves}
\affiliation{SLAC National Accelerator Laboratory,
Menlo Park, CA 94025, USA}

\author{Nima Arkani-Hamed}
\affiliation{Institute for Advanced Study, Princeton,
New Jersey 08540, USA}

\author{Sanjay Arora}
\affiliation{Dept.~of Physics and Astronomy, Rutgers University, Piscataway, NJ 08854, USA}

\author{Yang Bai}
\affiliation{SLAC National Accelerator Laboratory,
Menlo Park, CA 94025, USA}

\author{Matthew Baumgart}
\affiliation{Johns Hopkins University, Dept.~of Physics and Astronomy,
Baltimore, MD 21218, USA}

\author{Joshua Berger}
\affiliation{LEPP,
Cornell University, Ithaca, NY 14853, USA}

\author{Matthew Buckley}
\affiliation{Fermi National Accelerator Lab.,
Theory Group,
Batavia, IL 60510, USA }

\author{Bart Butler}
\affiliation{SLAC National Accelerator Laboratory, Menlo Park, CA 94025, USA}

\author{Spencer Chang}
\affiliation{University of Oregon,
Department of Physics,
Eugene, OR 97403-1274 USA}
\affiliation{University of California Davis,
Department of Physics,
Davis, CA 95616-8677, USA}

\author{Hsin-Chia Cheng}
\affiliation{University of California Davis,
Department of Physics,
Davis, CA 95616-8677, USA}

\author{Clifford Cheung}
\affiliation{Department of Physics, UC Berkeley, 
Berkeley CA, 94720, USA}

\author{R. Sekhar Chivukula}
\affiliation{Dept.~of Physics and Astronomy, Michigan State University
East Lansing, MI 48824, USA}

\author{Won Sang Cho}
\affiliation{IPMU, The University of Tokyo, Chiba, 277-8583, Japan}


\author{Randy Cotta}
\affiliation{SLAC National Accelerator Laboratory,
Menlo Park, CA 94025, USA}

\author{Mariarosaria D'Alfonso}
\affiliation{Dept.~of Physics, University of California Santa Barbara, Santa Barbara, CA 93106, USA}

\author{Sonia El Hedri}
\affiliation{SLAC National Accelerator Laboratory,
Menlo Park, CA 94025, USA}

\author{Rouven Essig \Ed}
\email{rouven@stanford.edu}
\affiliation{SLAC National Accelerator Laboratory,
Menlo Park, CA 94025, USA}

\author{Jared A.~Evans}
\affiliation{University of California Davis,
Department of Physics,
Davis, CA 95616-8677, USA}

\author{Liam Fitzpatrick}
\affiliation{Department of Physics, Boston University, 
Boston, MA 02215, USA}

\author{Patrick Fox}
\affiliation{Fermi National Accelerator Lab.,
Theory Group,
Batavia, IL 60510, USA }

\author{Roberto Franceschini}
\affiliation{EPFL, CH-1015 Lausanne, Switzerland}

\author{Ayres Freitas}
\affiliation{Dept.~of Physics and Astronomy, University of Pittsburgh,
Pittsburgh, PA 15260, USA}

\author{James S. Gainer}
\affiliation{High Energy Physics Division, Argonne National Laboratory, Argonne, IL
60439, USA}
\affiliation{Dept.~of Physics and Astronomy, Northwestern University,
Evanston, IL 60208, USA}

\author{Yuri Gershtein}
\affiliation{Dept.~of Physics and Astronomy, Rutgers University, Piscataway, NJ 08854, USA}

\author{Richard Gray}
\affiliation{Dept.~of Physics and Astronomy, Rutgers University, Piscataway, NJ 08854, USA}

\author{Thomas Gregoire}
\affiliation{Dept~of Physics, Carleton University, 
Ottawa, Ontario, K1S 5B6, Canada}

\author{Ben Gripaios}
\affiliation{CERN PH-TH, Case C01600, 1211 Geneva 23, Switzerland, USA}

\author{Jack Gunion}
\affiliation{University of California Davis,
Department of Physics,
Davis, CA 95616-8677, USA}

\author{Tao Han}
\affiliation{University of Wisconsin-Madison, Madison, WI 53706, USA}

\author{Andy Haas}
\affiliation{SLAC National Accelerator Laboratory,
Menlo Park, CA 94025, USA}

\author{Per Hansson}
\affiliation{SLAC National Accelerator Laboratory, Menlo Park, CA 94025, USA}

\author{JoAnne Hewett}
\affiliation{SLAC National Accelerator Laboratory,
Menlo Park, CA 94025, USA}

\author{Dmitry Hits}
\affiliation{Dept.~of Physics and Astronomy, Rutgers University, Piscataway, NJ 08854, USA}

\author{Jay Hubisz}
\affiliation{Department of Physics, Syracuse University, Syracuse, NY 13244, USA}

\author{Eder Izaguirre}
\affiliation{SLAC National Accelerator Laboratory,
Menlo Park, CA 94025, USA}

\author{Jared Kaplan}
\affiliation{SLAC National Accelerator Laboratory,
Menlo Park, CA 94025, USA}

\author{Emanuel Katz}
\affiliation{Department of Physics, Boston University, 
Boston, MA 02215, USA}

\author{Can Kilic}
\affiliation{Dept.~of Physics and Astronomy, Rutgers University, Piscataway, NJ 08854, USA}

\author{Hyung-Do Kim}
\affiliation{Dept.~of Physics and Astronomy, Seoul National University, Republic of Korea}

\author{Ryuichiro Kitano}
\affiliation{
Dept.~of Physics, Tohoku University,
Sendai 980-8578, Japan}

\author{Sue Ann Koay}
\affiliation{Dept.~of Physics, University of California Santa Barbara, Santa Barbara, CA 93106, USA}

\author{Pyungwon Ko}
\affiliation{Korea Institute for Advanced Study, Seoul 130-722, Republic of Korea}

\author{David Krohn}
\affiliation{Dept.~of Physics, Harvard University, 
Cambridge, MA 02138, USA}

\author{Eric Kuflik}
\affiliation{University of Michigan, Ann Arbor, Michigan 48109, USA}

\author{Ian Lewis}
\affiliation{University of Wisconsin-Madison, Madison, WI 53706, USA}

\author{Mariangela Lisanti \Ed}
\email{mlisanti@princeton.edu}
\affiliation{Princeton Center for Theoretical Science,
Princeton, NJ 08540, USA}

\author{Tao Liu}
\affiliation{Dept.~of Physics, University of California Santa Barbara, Santa Barbara, CA 93106, USA}

\author{Zhen Liu}
\affiliation{University of Wisconsin-Madison, Madison, WI 53706, USA}

\author{Ran Lu}
\affiliation{University of Michigan, Ann Arbor, Michigan 48109, USA}

\author{Markus Luty}
\affiliation{University of California Davis,
Department of Physics,
Davis, CA 95616-8677, USA}

\author{Patrick Meade}
\affiliation{YITP, State University of New York, Stony Brook, NY 11794, USA}

\author{David Morrissey}
\affiliation{TRIUMF, Vancouver, BC V6T 2A3, Canada}

\author{Stephen Mrenna}
\affiliation{Fermi National Accelerator Lab.,
Theory Group,
Batavia, IL 60510, USA }

\author{Mihoko Nojiri}
\affiliation{Theory Group, KEK, Ibaraki 305-0801, Japan}

\author{Takemichi Okui}
\affiliation{Dept.~of Physics, Florida State University, 
Tallahassee, FL 32306, USA}

\author{Sanjay Padhi}
\affiliation{University of California, San Diego, CA 92093, USA}

\author{Michele Papucci}
\affiliation{Theory Group, 
Lawrence Berkeley National Laboratory
Berkeley, CA 94720, USA}

\author{Michael Park}
\affiliation{Dept.~of Physics and Astronomy, Rutgers University, Piscataway, NJ 08854, USA}

\author{Myeonghun Park}
\affiliation{Dept.~of Physics, University of Florida, Gainesville, FL 32611, USA}

\author{Maxim Perelstein}
\affiliation{LEPP,
Cornell University, Ithaca, NY 14853, USA}

\author{Michael Peskin}
\affiliation{SLAC National Accelerator Laboratory,
Menlo Park, CA 94025, USA}

\author{Daniel Phalen}
\affiliation{University of California Davis,
Department of Physics,
Davis, CA 95616-8677, USA}

\author{Keith Rehermann}
\affiliation{CTP, 
Massachusetts Institute of Technology, 
Cambridge, MA 02139, USA}

\author{Vikram Rentala}
\affiliation{University of California, Irvine
Irvine, CA 92697, USA}

\author{Tuhin Roy}
\affiliation{Department of Physics, University of Washington, Seattle, WA 98195-1560, USA}

\author{Joshua T. Ruderman}
\affiliation{Dept.~of Physics, Princeton University, Princeton, NJ 08542, USA}

\author{Veronica Sanz}
\affiliation{York University,Toronto, Ontario, M3J 1P3, Canada}

\author{Martin Schmaltz}
\affiliation{Department of Physics, Boston University, 
Boston, MA 02215, USA}

\author{Stephen Schnetzer}
\affiliation{Dept.~of Physics and Astronomy, Rutgers University, Piscataway, NJ 08854, USA}

\author{Philip Schuster \Ed}
\email{pschuster@perimeterinstitute.ca}
\affiliation{Perimeter Institute for Theoretical Physics,
Ontario, Canada, N2L 2Y5 }
\affiliation{Institute for Advanced Study, Princeton,
New Jersey 08540, USA}

\author{Pedro Schwaller}
\affiliation{Physics Department, University of Illinois at Chicago, Chicago, IL 60607, USA}
\affiliation{High Energy Physics Division, Argonne National Laboratory, Argonne, IL
60439, USA}
\affiliation{Universi\"at Z\"urich, 
Institut f\"ur Theoretische Physik, 8057 Z\"urich, Switzerland}

\author{Matthew D.~Schwartz}
\affiliation{Dept.~of Physics, Harvard University, 
Cambridge, MA 02138, USA}

\author{Ariel Schwartzman}
\affiliation{SLAC National Accelerator Laboratory,
Menlo Park, CA 94025, USA}

\author{Jing Shao}
\affiliation{Department of Physics, Syracuse University, Syracuse, NY 13244 USA}

\author{Jessie Shelton}
\affiliation{Yale University, Sloane Physics Lab, New Haven, CT 06520, USA}

\author{David Shih}
\affiliation{Dept.~of Physics and Astronomy, Rutgers University, Piscataway, NJ 08854, USA}

\author{Jing Shu}
\affiliation{IPMU, The University of Tokyo, Chiba, 277-8583, Japan}

\author{Daniel Silverstein}
\affiliation{SLAC National Accelerator Laboratory, Menlo Park, CA 94025, USA}

\author{Elizabeth Simmons}
\affiliation{Dept.~of Physics and Astronomy, Michigan State University
East Lansing, MI 48824, USA}

\author{Sunil Somalwar}
\affiliation{Dept.~of Physics and Astronomy, Rutgers University, Piscataway, NJ 08854, USA}

\author{Michael Spannowsky}
\affiliation{University of Oregon,
Department of Physics,
Eugene, OR 97403-1274 USA}

\author{Christian Spethmann}
\affiliation{Department of Physics, Boston University, 
Boston, MA 02215, USA}

\author{Matthew Strassler}
\affiliation{Dept.~of Physics and Astronomy, Rutgers University, Piscataway, NJ 08854, USA}

\author{Shufang Su}
\affiliation{Dept.~of Physics, University of Arizona, Tucson, AZ 85721, USA}
\affiliation{University of California, Irvine
Irvine, CA 92697, USA}

\author{Tim Tait \Ed}
\email{ttait@uci.edu}
\affiliation{University of California, Irvine
Irvine, CA 92697, USA}

\author{Brooks Thomas}
\affiliation{University of Hawaii, Honolulu, HI 96822, USA}

\author{Scott Thomas}
\affiliation{Dept.~of Physics and Astronomy, Rutgers University, Piscataway, NJ 08854, USA}

\author{Natalia Toro \Ed}
\email{ntoro@perimeterinstitute.ca}
\affiliation{Perimeter Institute for Theoretical Physics,
Ontario, Canada, N2L 2Y5 }
\affiliation{Institute for Advanced Study, Princeton,
New Jersey 08540, USA}

\author{Tomer Volansky}
\affiliation{Department of Physics, UC Berkeley, 
Berkeley CA, 94720, USA}

\author{Jay Wacker \Ed}
\email{jgwacker@stanford.edu}
\affiliation{SLAC National Accelerator Laboratory,
Menlo Park, CA 94025, USA}

\author{Wolfgang Waltenberger}
\affiliation{Institut f\"ur Hochenergiephysik, Vienna, Austria}

\author{Itay Yavin}
\affiliation{	CCPP, New York University, New York, NY 10003, USA}

\author{Felix Yu}
\affiliation{University of California, Irvine
Irvine, CA 92697, USA}

\author{Yue Zhao}
\affiliation{Dept.~of Physics and Astronomy, Rutgers University, Piscataway, NJ 08854, USA}

\author{Kathryn Zurek}
\affiliation{University of Michigan, Ann Arbor, Michigan 48109, USA}

\collaboration{LHC New Physics Working Group}

%% file: abstract.tex
\date{\today}
\begin{abstract}
\newpage
This document proposes a collection of simplified models relevant to the design of new-physics searches at the LHC and the characterization of their results.  
Both ATLAS and CMS have already presented some results in terms of simplified models, and we encourage them to continue and expand this effort, which supplements both signature-based results and benchmark model interpretations.  
A simplified model is defined by an effective Lagrangian describing the interactions of a small number of new particles.
Simplified models can equally well be described by a small number of masses and cross-sections.  These parameters are directly related to collider physics observables, making simplified models a particularly effective framework for evaluating searches and a useful starting point for characterizing positive signals of new physics. 
This document serves as an official summary of the results from the ``Topologies for Early LHC Searches'' workshop, held at SLAC in September of 2010, the purpose of which was to develop a set of representative models that can be used to cover all relevant phase space in experimental searches.
Particular emphasis is placed on searches relevant for the first $\sim 50 - 500$ pb$^{-1}$ of data and those motivated by supersymmetric models.  
This note largely summarizes material posted at \texttt{http://lhcnewphysics.org/}, which includes simplified model definitions, Monte Carlo material, and supporting contacts within the theory community. 
We also comment on future developments that may be useful as more data is gathered and analyzed by the experiments.
\end{abstract}

%% file: introduction.tex
\section{Introduction}

After decades of preparation, the experiments at the Large Hadron Collider (LHC) are taking the first steps toward resolving many long-standing puzzles about fundamental physics at the weak scale.
With the first $\sim 40$ pb$^{-1}$ of data obtained in 2010, both ATLAS and CMS are already probing new physics beyond the reach of any past experiment.
Two principal questions should be addressed with these early LHC searches.  
The first is whether certain classes of new physics can evade the existing ATLAS and CMS search programs, but still be detectable with new search techniques or optimization strategies.  Second, it is important to understand the physical implications of new-physics searches, whether they see evidence for new physics or constrain it.  We propose that very simple models of new physics, involving relatively few particles and decay modes, offer a natural framework for both tasks.  ATLAS and CMS could enhance the applicability of new-physics searches by considering 
their sensitivity to such ``simplified models''.  In this document, we discuss an illustrative example and present a catalog of representative models to cover a broad space of signatures with a broad range of new-physics models.  

Recent efforts, including two joint experiment-theory workshops at CERN \cite{JuneWorkshop, NovWorkshop}  have focused on using pre-defined simplified models in the design of new-physics searches and characterization of their results.  Indeed, the results of some of the first searches for supersymmetry at ATLAS and CMS have been represented in terms of simplified models, see for example \cite{Collaboration:2011qk, Aad:2011ks, Collaboration:2011xk, Aad:2011xm, ATLAS-CONF-2011-039, Collaboration:2011wc, CMS-PAS-SUS-10-009,CMS-PAS-SUS-11-001,CMS-PAS-SUS-10-005}.  To complement this effort, a workshop was held at SLAC to define a set of simplified models whose topologies are representative of the wide variety of new-physics possibilities that could be seen at the LHC \cite{SLACWorkshop}.    This document is largely a catalog of the simplified models developed at the SLAC workshop.  

Our hope is that the simplified models listed here will provide a foundation for assessing the impact of existing searches, and how they can be extended or better optimized.  In addition, we expect that the simplified models here will be a useful starting point for characterizing any evidence for new physics, in a systematic and unbiased manner.

\subsection{The Purpose of Simplified Models}
A model of new physics is defined by a TeV-scale effective Lagrangian describing its particle content and interactions. A simplified model is specifically designed to involve only a few new particles and interactions.  Many simplified models are limits of more general new-physics scenarios, where all but a few particles are integrated out.   Simplified models can equally well be described by a small number of parameters directly related to collider physics observables: particle masses (and their decay widths, which can sometimes be neglected), production cross-sections, and 
branching fractions.  

Simplified models are clearly not model-independent, but they do avoid some pitfalls of model-dependence.  The sensitivity of any new-physics search to a few-parameter simplified model can be studied and presented as a function of these parameters and in particular over the full range of new particle masses.  Though defined within a simplified model, these topology-based limits also apply to more general models giving rise to the same topologies. 

The primary intended applications for simplified model results are as follows:
\begin{itemize}
\item {\bf Identifying the boundaries of search sensitivity:}
Any critical assessment of LHC searches needs to include a clear identification 
of the boundaries of sensitivity  --- for example, the dependence of reconstruction and selection efficiencies on the mass differences between a parent particle and its decay products.   One- and two-dimensional slices within a simplified model can illustrate these boundaries very clearly.  Only with this information can experimentalists and theorists identify kinematic ranges (or entire topologies!) for which existing search strategies are not efficient, and devise appropriate generalizations to these strategies.  
For the same reasons, limits on simplified models also serve as a valuable reference for theorists who wish to estimate a search's sensitivity to alternative new-physics models in their own Monte Carlo.
 
\item {\bf Characterizing new physics signals:}
If new physics is observed, it will be important to fully characterize the range
of particle quantum numbers, masses, and decay topologies that it may involve. 
As has been discussed in \cite{Alwall:2008ag}, simplified models can offer a natural starting point 
for quantifying the consistency of a signal with different kinds of physics reactions.  Similar strategies have been discussed in \cite{Knuteson:2006ha,ArkaniHamed:2007fw}. 

\item {\bf Deriving limits on more general models:}
Constraints on a wide variety of models can be deduced from limits on simplified models.  Within each final state, simplified model limits can be formulated as an upper limit on the number of events in a signal region, and a parametrized efficiency for each simplified-model topology to populate the signal region.  Limits on other models giving rise to the same topologies can be inferred by summing the effective cross-section for each topology (a product of cross-sections and branching ratios),  weighted by their experimental efficiencies, and comparing the result to the upper bound.  This procedure can be extended to multiple signal regions if a combined likelihood is reported as a function of the number of signal events in each signal region.  These procedures are discussed in several talks at the 
workshop \cite{NovWorkshop, DavisWorkshop} and, for example, in \cite{Alves:2011sq}.  We also give an example in  
Section \ref{sec:topologies}.  It should be emphasized that this procedure yields weaker limits than the direct study of experimental efficiencies for a given specific model, as the procedure uses only  topologies populated by \emph{both}  the specific \emph{and} simplified models.  This procedure should therefore be regarded as an initial check only, which can be followed by a dedicated study or RECAST-style analysis \cite{Cranmer:2010hk} if higher precision is needed.  
\end{itemize}
Finally, we note that simplified models can be simulated either as modules from widely used model frameworks (like the MSSM) in Pythia~\cite{Sjostrand:2006za} or MadGraph~\cite{Alwall:2007st}, as new models in MadGraph, or as OSETs using Marmoset  \cite{ArkaniHamed:2007fw} or recent versions of Pythia. 

\subsection{Scope and Future Developments}
While the set of simplified models described in this note is extensive, it is also incomplete.  Some omissions are by design -- we have tried to avoid duplicate structure by focusing on simple representative topologies, even when they have straightforward extensions with, e.g., longer decay chains.  With this exception, we have endeavoured to produce a thorough catalog of simplified models.  
However, good reference models are certainly missing; we invite readers 
to point these out, and to contribute new simplified models to the web database at \verb=http://lhcnewphysics.org/=. 

Many individuals involved in new-physics searches in ATLAS and CMS have requested a short list of ``high priority'' simplified models.
A first attempt at this task, focusing on simplified models appropriate for `canonical' supersymmetry searches, was
undertaken at a CERN workshop \cite{NovWorkshop}, but only limited consensus was reached.  Prioritizing simplified models is beyond the scope of this document.  However, it is clear that as some simplified models are incorporated in ATLAS and CMS searches, it will be valuable to further standardize conventions for particular simplified models, such as reference cross-sections, parameter slices of interest, and even shared Les Houches Accord spectra \cite{Allanach:2008qq} that are publicly available and shared between the two experiments. Such efforts would best be coordinated by a joint ATLAS/CMS/theory working group.

\subsection{Using This Note}
Section \ref{detailed} describes one simplified model example in detail, illustrating the procedure of parameterizing the model and applying it to search results.  Following this discussion, the main body of the text is a catalog of brief definitions of simplified models.  Simplified models are organized according to classes of signatures --- those involving jets (\S \ref{jets}), heavy-flavor ($b$ or $\tau$, \S \ref{heavy}), leptons (\S \ref{leptons}), photons (\S \ref{photons}), and `exotic objects' such as new displaced vertices, non-standard timing, or novel jet-like structures (\S \ref{exotica}).  Multi-object signatures are grouped according to the \emph{last} category they populate.  For example, fully leptonic and semi-leptonic top decays are classified as leptonic signatures, hadronic top decays are classified as heavy-flavor, and any model with new displaced vertices is classified as `exotic'.  Of course, most simplified models populate multiple signatures.  In this case, we have placed each simplified model under the signature that would likely be the dominant discovery mode for that model.

Most of the simplified models described in this document are discussed in more detail in write-ups  at \texttt{http://lhcnewphysics.org/} \cite{LHCNPWG}.  Registration is open, and registered users can upload new simplified models and post comments.  Experimentalists should feel free to use these postings to contact the authors of the simplified models.  In most cases, the online write-ups provide more detailed definitions of the simplified models as well as suggestions for simulating, parametrizing, and searching for each simplified model.

%% file: longexamples.tex
\section{A Detailed Example of a Simplified Model} \label{detailed}

This section, adapted from~\cite{Alves:2011sq},  outlines the important elements that go into any simplified model analysis.  As an illustrative example,  it focuses on gluino production and decay as a model for hadronic jets plus missing energy signals.  We will discuss how limits can be set in a multidimensional parameter space and how the limits from multiple topologies can be combined.  The procedure outlined here is a general one and can be applied to any of the simplified models listed in this review.

\subsubsection{Effective Lagrangian}

Consider a direct three-body gluino decay into an electroweak gaugino and two light-flavored quarks,
\begin{center}{
\large $\tilde{g} \rightarrow q \bar{q}^\prime \chi^0$}.
\end{center}
This decay mode occurs in supersymmetric models where the squarks are significantly heavier than the gluino; it 
proceeds through the dimension-six operator
\begin{eqnarray}
\LL_{\text{int}} =   \frac{\lambda_i^2}{M_{i}^2} {\tilde{g}} q_i \bar{q}_i{\chi^0} +\hc ,  
\end{eqnarray}
where $i$ runs over the different quark flavors, $\lambda_{i}$ is the Yukawa coupling for the  quark-squark-${\chi^0}$  vertex, and $M_i$ is the effective scale of the interaction.   The flavor structure of the final state is determined by the mass spectrum of the corresponding squarks, with decays through lighter mass squarks occurring more rapidly.  In this example, only light-flavor decay modes are considered (see \S\ref{heavy:gluino} for the analogous heavy-flavor discussion). 
\begin{figure}[b]
\begin{center}
\includegraphics[width=6in]{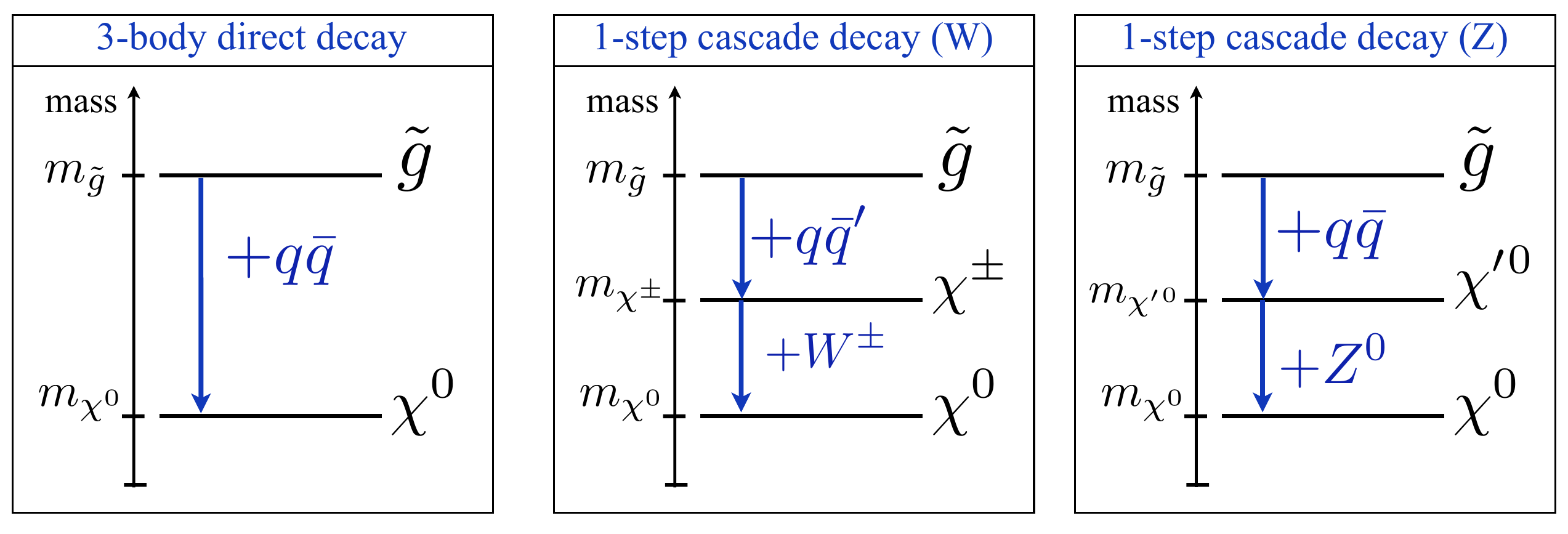}
 \end{center}
 \caption{Illustrations of the three gluino simplified models discussed in this section.
  \label{Fig: Spectra}
 }
 \end{figure}

Direct three-body decays arise in models where the squarks are decoupled, such as in split-supersymmetry \cite{Kilian:2004uj}, or where the soft masses of the squarks are at the TeV-scale, but are still somewhat larger than the gluino mass.  These decays dominate when
\begin{itemize}
\item ${\chi^0} = \widetilde{B}$  and the right-handed squarks are lightest, or the $\widetilde{W}$ is kinematically inaccessible
\item ${\chi^0} = \widetilde{W}$ and the left-handed squarks are lightest, or all squark masses are comparable
\item ${\chi^0} = \widetilde{H}$ and the heavy-flavor squarks are kinematically accessible in gluino decays, or the $\widetilde{B}$ and $\widetilde{W}$ are kinematically inaccessible.
\end{itemize}
In mSUGRA \cite{Ibanez:1981yh,Chamseddine:1982jx,Barbieri:1982eh,Ohta:1982wn,Hall:1983iz} 
and GMSB-like \cite{Dine:1981za,Dimopoulos:1981au,Dine:1981gu,Dine:1982zb,Nappi:1982hm,AlvarezGaume:1981wy,Dimopoulos:1982gm,Dine:1993yw,Dine:1994vc,Dine:1995ag,Giudice:1998bp}
models, the LSP is usually bino-like and there is no strong splitting between the left and right-handed squarks; 
therefore, the direct decays usually do not dominate.  
In contrast, AMSB scenarios \cite{Randall:1998uk,Giudice:1998xp,Pomarol:1999ie,Gherghetta:1999sw}
have a wino-like LSP and a large wino gauge-Yukawa coupling, leading to a large branching ratio for three-body gluino decays.

A complementary simplified model corresponds to the case where the gluino goes through a three-body decay to a chargino that subsequently decays to a gauge boson and the LSP,
\begin{center}{
\large ${\tilde{g}}\rightarrow q \bar{q}^\prime {\chi}^\pm \rightarrow q \bar{q}^\prime (W^\pm {\chi^0})$ \qquad or \qquad
${\tilde{g}}\rightarrow q \bar{q} {\chi}'{}^0 \rightarrow q \bar{q} (Z^0 {\chi^0})$
}
\end{center}
The decay chain \mbox{``gluino $\rightarrow$ heavy electroweakino $\rightarrow$ lightest electroweakino"} is preferred in many supersymmetric scenarios \cite{Barnett:1987kn}, including mSUGRA.  A similar chain \mbox{``KK-gluon $\rightarrow$ KK-gauge boson $\rightarrow$ KK-graviton"} is also present in Extra Dimensions \cite{Appelquist:2000nn,Dienes:1998vh,Cheng:2002ab}.

When the intermediate particle is a chargino, all events have two $W^\pm$ bosons in the final state. 
Alternative simplified models exist in which the intermediate state is  neutral and decays to a $Z^0$ boson or higgs instead of a $W^\pm$. 
When exchanging a $W^\pm$ for a $Z^0$, the mass difference is a small effect at the LHC.  However, the difference between their leptonic decay modes is quite significant.  In hadronic searches, the difference between modes is manifested in two ways: the fraction of events that are truly hadronic, and the presence in the $W$ mode of leptonic $W$'s that are not vetoed in the searches (e.g., if the lepton is non-isolated or out of acceptance).  These effects are unlikely to affect the optimization of search regions, but do introduce complications in translating limits from one simplified model to another.   Answering this question requires understanding the differences in the acceptances/efficiencies for events with $Z^0$-final states versus $W^\pm$-final states.   

\subsubsection{Simplified Model Parametrization}

A simplified model is described by a minimal set of parameters that often include the particle masses and the production cross sections.  For example, the three-body direct decay model is parametrized in terms of $m_{\tilde{g}}, m_{\tilde{\chi}^0}, \text{and } \sigma(pp\rightarrow {\tilde{g}}{\tilde{g}}+X)$.  The one-step cascade decay introduces two new parameters: the mass of the intermediate particle $m_{\chi^\pm}$ and the branching ratio of ${\tilde{g}}$ decaying into $\chi^\pm$.  However, it is much easier to consider each simplified model with branching ratios set to 100\%.  Models with multiple decay modes can be studied by taking linear combinations of results for single decay modes, as discussed in the following section (\ref{sec:topologies}).   When the efficiencies of a search for two decay modes are very different, studies of `mixed' topologies may also be desirable.  

Assuming a 100\% branching ratio reduces the number of parameters in the one-step cascade model to four.   The choice of $m_{\chi^\pm}$ alters the kinematics of the theory and must be included, despite the challenges of presenting limits in a four-dimensional space.  It is instructive to consider lower-dimensional ``mass slices'' in $m_{\chi^\pm}$, which illustrate the distinctive features of the one-step cascade and capture all the relevant corners of phase space.  An example of a useful family of chargino mass slices is
\begin{equation}
\label{Eq: 1stepSlicings}
m_{{\chi}^\pm}=m_{{\chi^0}}+r(m_{{\tilde{g}}}-m_{{\chi^0}}).
\end{equation}
The case of $r=0$ is identical to the direct three-body decay.  The case of $r=1$ closely resembles a direct two-body gluino decay, provided the $W^\pm$ is boosted so that its decay products merge together.  A few intermediate values of $r$ (e.g. 0.25, 0.5, and 0.75) cover a variety of kinematics.  In hadronic searches, the limit of small $r$ approaches the direct three-body decay, but the precise  $\chi^\pm - \chi^0$ mass difference significantly affects the sensitivity of leptonic searches.  For these, a mass slice with $m_{\tilde g}$ fixed near the limit of detectability, and $m_{\chi^0}$ and $m_{\chi^\pm}$ varied independently, is also relevant.

To explore the effect of on-shell decays near threshold, the alternative mass slice 
\begin{eqnarray}
\label{Eq: CharginoThreshold}
m_{{\chi}^\pm}\simeq m_{{\chi^0}}+ m_{W^\pm},
\end{eqnarray}
is useful.  Threshold effects are fairly modest because the mass scales accessible at the LHC are sufficiently above $m_{W^\pm}$, though they do become important for lighter gluino masses.  In \cite{Alves:2010za}, this can be seen as a sharp drop in the cross section sensitivity along the line in Eq.~\ref{Eq: CharginoThreshold}.

\subsubsection{Combining Topologies}\label{sec:topologies}
\begin{figure}[tb]
\begin{center}
\includegraphics[width=4in]{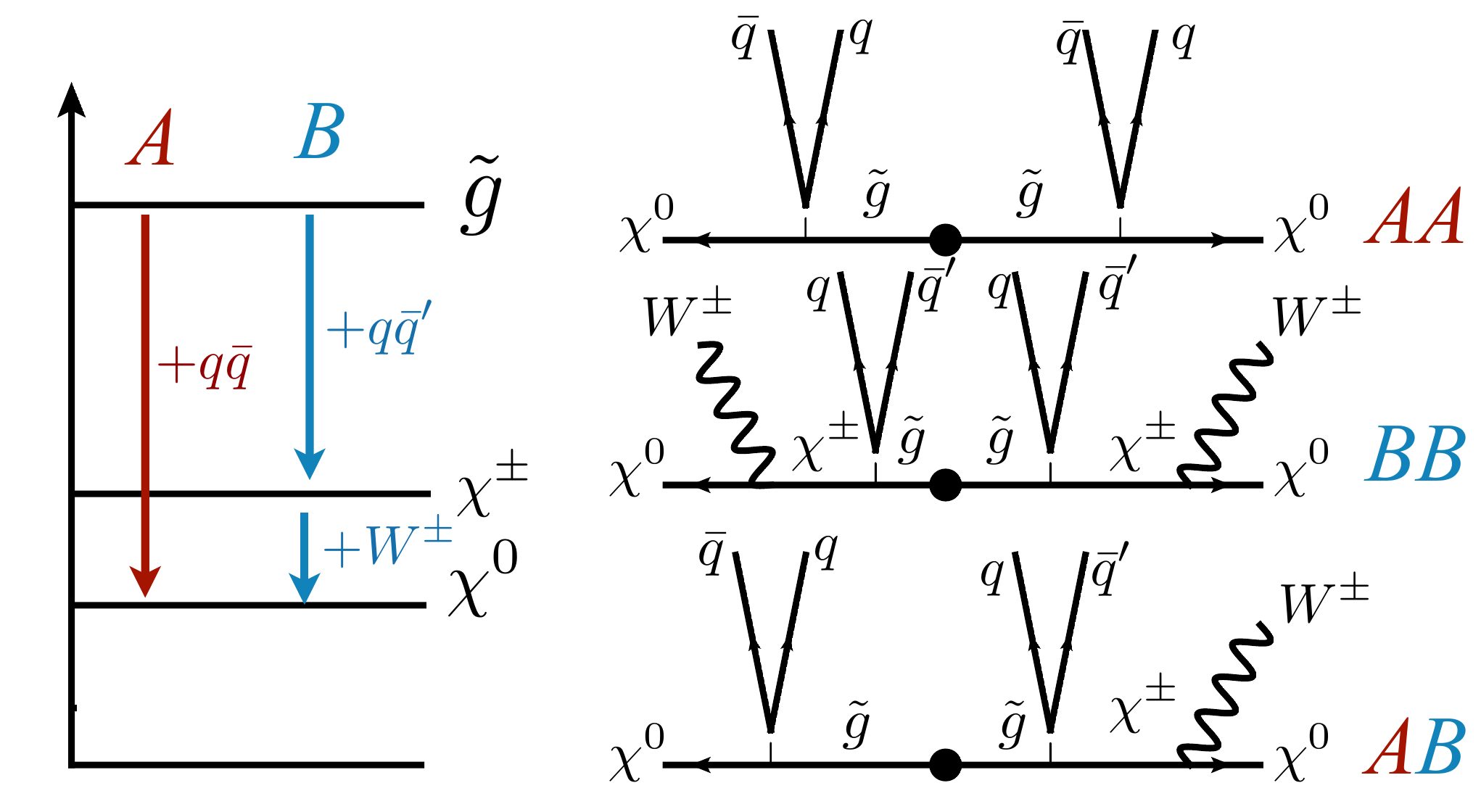}
 \end{center}
 \caption{
 \label{Fig: HybridFig}
$A=$ three-body direct decay 
 and $B=$ one-step cascade decay. Three topologies are possible: symmetric $AA$, $BB$ and mixed $AB$.  
 }
 \end{figure}

The above discussion has focused on topologies corresponding to particle-antiparticle pair production, with the two produced particles decaying through identical channels.  More generally, associated production topologies  and `mixed' decay modes (where, for instance, one gluino decays directly to the LSP (mode $A$) while the other decays through a cascade (mode $B$)), as in the lower diagram of Figure \ref{Fig: HybridFig}.  

It is useful to consider what one may infer about these models given \emph{only} a search's sensitivity to the two `symmetric'  decay modes.  We consider this question in the context of an idealized search result with two components: an upper limit $N_{max}$ on the expected number of signal events in a signal region of interest, and the efficiency $\epsilon$ for each process to populate this signal region.   If all efficiencies were known, one could infer a cross-section limit $\sigma_{max}$ for models with branching ratios $B_{A}$, $B_{B}$ by 
\begin{equation}
\sigma_{max} = \frac{N_{max}}{B_{A}^2 \epsilon_{AA} + 2 B_{A} B_{B} \epsilon_{AB} + B_{B}^2 \epsilon_{BB}}\label{masterSigma}
\end{equation}
(the cross-section upper limits for the two symmetric decays are simply $\sigma_{max,AA} = N_{max}/\epsilon_{AA}$ and similarly for the mode $BB$).  However, we wish to consider what can be gleaned about  $\sigma_{max}$  if the efficiency $\epsilon_{AB}$ is unknown.  

Upper and lower bounds on $\sigma_{max}$ can be obtained simply by using the fact that $0 \leq \epsilon_{AB} \leq 1$.  The lower bound corresponding to $\epsilon_{AB}\rightarrow 0$ amounts to ``throwing out'' the mixed events.  The resulting limit is conservative (it always under-estimates the true strength of a search result),  but can be a considerable underestimate of the actual search sensitivity, particularly when both branching ratios are comparable or the dominant decay mode has low efficiency.  

In many cases where the decay modes $A$ and $B$ produce similar final states, the mixed decay modes have an efficiency comparable to those of the two symmetric modes, and typically intermediate:
\begin{equation}
\min\left(\epsilon_{AA}, \epsilon_{BB}\right) \leq \epsilon_{AB} \leq \max\left(\epsilon_{AA}, \epsilon_{BB}\right).\label{tightInequality}
\end{equation}
If the $\epsilon_{AA}$ and $\epsilon_{BB}$ are comparable, then inserting these bounding values into \eqref{masterSigma} allows a fairly precise determination of $\sigma_{max}$, even when branching ratios are nearly evenly split between the two decay modes.   

It is important to emphasize that \eqref{tightInequality} is by no means guaranteed.  When expected violations of \eqref{tightInequality} are large, the mixed topologies warrant careful dedicated study.   As an extreme example, if mode $A$ is fully hadronic and mode $B$ typically produces a lepton, then for a one-lepton search one expects  $\epsilon_{AB} \gg \epsilon_{AA}, \epsilon_{BB}$.   These correspond to cases where it is clearly important to parametrize a search's sensitivity to the mixed decay modes directly.  However, in the case of hadronic searches and the gluino decays shown in Fig. \ref{Fig: HybridFig}, \eqref{tightInequality} is typically true at least to a good approximation.  A reasonable assessment of whether \eqref{tightInequality} is likely to hold can be obtained by studying the step-by-step efficiencies of a search for the two symmetric decay modes.  If mode $AA$ passes each individual cut with comparable or greater efficiency than $BB$ (or vice versa), then \eqref{tightInequality} is likely to hold.  Even when this is not the case, the \emph{lower} bound is robust in many examples.  Thus, in most cases one may draw powerful conclusions from the symmetric decay modes alone.

%% file: jets.tex
\section{Jets}\label{jets}

Jet-based simplified models are focused exclusively on hadronic channels with and without missing energy.   
The channels are organized based upon the number of final state jets and the production mode (i.e., through a resonance or two-to-two scattering).
\begin{table}[b]
\begin{tabular}{|c||c|c||c|c|c|}
\hline
$N_j$ &
\multicolumn{2}{|c||}{MET}&
\multicolumn{3}{|c|}{NoMET}\\
\hline
& $2\rightarrow 1$ &$2\rightarrow 2$& $2\rightarrow 1$ &$2\rightarrow 2$&$2\rightarrow 3$\\
\hline
1&\ref{J.000}\quad *& \ref{J.005}\quad *&--&--&--\\
& {\tt J.000}& {\tt J.005} & &&  \\
\hline
2&\ref{J.001}&\ref{J.006}, \ref{J.009}&\ref{J.012}.&*&--\\
&{\tt J.001}&{\tt J.006}, {\tt J.009} &{\tt J.012}&  &\\
\hline
3& \ref{J.001}&\ref{J.007}&&&\ref{J.011} \\
&{\tt J.001}&{\tt J.007}&&& {\tt J.011}\\
\hline
$4^+$&\ref{J.001}&*&\ref{J.004}&\ref{J.003}&--\\
&{\tt J.001}&&{\tt J.004}&{\tt J.003}&\\
\hline
\end{tabular}
\caption{\label{Tab: JetSigs} Jet-based simplified model organization. References to the relevant section in this note are given (e.g. \ref{J.000}), as well
as posting numbers at www.lhcnewphysics.org (e.g. J.001).  
$*$ indicates that there are additional models that do not have write ups currently available on {\tt http://lhcnewphysics.org}.}
\end{table}
Links to the available supporting information at www.lhcnewphysics.org are given by the post number (e.g. J.011). 

\subsection{Three Jets and No $\MET$ from an Anomalous Three Gluon Coupling}
\label{J.011}
This model explores three jet events without 
$\MET$ that arise from a unique gauge invariant, CP-even
operator that couples three gluon fields, and which cannot be
re-expressed as a four-quark contact interaction.  This operator can
be induced by new heavy colored states in loops or by gluon
compositeness.  In addition to the three-jet topology, this operator can change the total cross section and angular distributions of $t\bar{t}$ production relative to the Standard Model.  While existing limits are weak, searches for these signatures at the LHC should be able to place an upper limit on the coefficient
of this effective operator. 
A complete discussion of the signatures of as well as the existing
and prospective limits on this operator can be found
in~\cite{Simmons:1989zs, Simmons:1990dh, Dreiner:1991xi, Cho:1993eu,
Cho:1994yu, Zhang:2010px, Zhang:2010dr, Duff:1991ad, Dixon:1993xd}.\\
\authorcite{J.011}{J. Gainer, M. Schwartz}

\subsection{Multijets from an Initial Resonance (no $\MET$)}
\label{J.004}
This simplified model consists of two new particles: the R particle can be resonantly produced at a hadron collider, either from quarks or gluons in the initial state, and decays predominantly to a pair of secondary resonances P, which themselves decay to two jets (quarks or gluons).  The P particles can be pair-produced from minimal QCD interactions as well as from R decays.  By assumption, the mass of the R particle must be greater than twice the P mass.  Achieving a sizable production rate for R requires an order one coupling to either quarks or gluons, which means that a substantial coupling R-P-P is needed in order for the dominant decay channel of R to be a pair of P particles.  This suggests that R can be a broad resonance.  Because the branching fraction of R to dijets is small compared to the dominant decay channel, dijet resonance searches for R can be avoided.  On the other hand, in order for P to be consistent with dijet resonance constraints, the coupling that gives rise to its decay to dijets must be small, which implies that the intrinsic width of P must be small. If R is produced from a quark/antiquark initial state, then R decays may be the dominant source of P pair production at the Tevatron, while at the LHC QCD production may dominate the cross section due to the difference between gluon and quark PDFs.  In either case, the dominant signal is the production of four hard jets that can be paired up to discover the P resonances.  If R decays are the dominant source of P production, then the four-jet invariant mass in signal events will also reveal the presence of R.  A
concrete realization of such a model is provided by vectorlike confining theories 
\cite{Kilic:2009mi} at the TeV scale that naturally avoid low energy precision constraints from flavor and oblique electroweak corrections.  
In such models, the rho meson of the new strong interaction plays the role of R while the pions play the role of P.  The discovery potential for these models has been studied for the Tevatron and the LHC in \cite{Kilic:2010et,Kilic:2008ub,Kilic:2008pm} and was found to be promising.\\
\authorcite{J.004}{D. Alves, C. Kilic}

\subsection{Multijets from Pair-Production (no $\MET$)}
\label{J.003}
This simplified model introduces two new particles, S and G, both of which are colored and can be pair-produced at hadron colliders, while resonant production of either one is suppressed or absent.  Beyond the minimal QCD interactions, there are two new couplings. The first one couples S-G-q, where q is a quark, while the second couples S-q-q.  The second coupling must be small in order to avoid dijet resonance constraints, while the first coupling is less constrained.  If S is lighter than G, then it decays directly to dijets while G decays to three jets through an intermediate S resonance.  If G is lighter, then it will decay to three jets through an off-shell S, while an S in this case will decay to four jets through an intermediate G resonance.  Depending on the mass hierarchy and the dominant production mechanism, the final state can contain four to eight hard jets.  Both S and G can be narrow resonances. This scenario can be realized in supersymmetric models with R-parity violation (through the UDD term) \cite{Barger:1989rk,Allanach:1999ic,Dreiner:1997uz,Dreiner:2006gu,Dreiner:1991pe}
and with heavy charginos and neutralinos, where S is a squark that decays through R-parity violating couplings and G is the gluino.
Signatures of this simplified model have been studied 
in \cite{Chivukula:1991zk,Choudhury:2005dg,Dobrescu:2007yp,Plehn:2008ae}. \\
\authorcite{J.003}{D. Alves, C. Kilic}

\subsection{Two jets and $\MET$ from Quark Partners}
\label{J.009}

This model consists of quark, lepton, and gauge boson partners that have the same spin as their corresponding Standard Model (SM) counterparts and is expanded in detail in Ref. \cite{Nojiri:2011qn}.  Gauge couplings of the gauge boson partners are the same as the corresponding SM ones.  The masses of the new quark and gauge boson partner states are free parameters.  This model is a simple extension of the LHT \cite{ArkaniHamed:2001nc,Schmaltz:2005ky,Katz:2003sn,Batra:2004ah,Csaki:2005fc,Batra:2007iz,ArkaniHamed:2002pa,Perelstein:2003wd,ArkaniHamed:2002qx,ArkaniHamed:2002qy,Kaplan:2003uc,Schmaltz:2004de,Gregoire:2002ra,Low:2002ws,Skiba:2003yf} 
or UED \cite{Appelquist:2000nn,Dienes:1998vh,Cheng:2002ab}, 
and may be used to study the spin dependence of the distributions of  SUSY-like signatures.  As an example, the two jets and $\MET$ signature is considered.  The quark partners are likely polarized when produced at the LHC, and this effect may be observed in ETmiss/Meff distributions.  When the average polarization is small due to the interference of the gauge boson exchanges, the azimuthal angle correlation of the leading jets from the quark partner decay is sensitive to the spin structure.  These results suggest that the LHC can distinguish the spin of quark partners in the 2 jets + $\MET$ +X channel.\\
\authorcite{J.009}{M. Nojiri, J. Shu}

 \subsection{Dijets from Colored Resonances (no $\MET$)}
 \label{J.012}
This simplified model studies colored resonance production at the LHC.  The possible colored resonances are classified based on group theory decomposition, and their effective interactions with light partons are then determined. 
The production cross section from annihilation of valence quarks or gluons can be on the order of  $400-1000$ pb at LHC energies for a mass of 1 TeV with nominal couplings, leading to the largest production rates for new physics at the TeV scale,
and the simplest event topology with dijet final states.
This simplified model formulation is readily applicable to future searches with other decay modes.
A complete discussion of the interactions, including LHC production rates, current bounds, and a full set of references, can be found in \cite{Han:2010rf}.\\
\authorcite{J.012}{T. Han, I. Lewis, Z. Liu}
 
 \subsection{Multijets+$\MET$ from an Initial Resonance}
 \label{J.001}
 
There are three variants of this simplified model, which covers topologies consisting of multijets (2-4 partons) and missing energy arising from the cascade decay of an initial resonantly produced state.  One variant is R-parity conserving SUSY with a resonance (motivated by e.g. high scale 
physics \cite{Cleaver:1997jb,Langacker:1998tc} or
phenomenology \cite{Cvetic:1997ky,Batra:2003nj,Kang:2009rd,Langacker:2007ac,Barger:2004bz,Kang:2004pp});
 the other two variants model cases where intermediate resonances can be constructed,
and differ from each other by the color representation of the initial resonance. The models are modularly extendable and designed to be
simply implemented in \texttt{MadGraph} or similar Monte Carlo programs.  Two interesting kinematic regions are identified and relevant search strategies are discussed.  \\
\authorcite{J.001}{J. Shelton, M. Spannowsky}

\subsection{Two or More Jets + $\MET$ from Squark Production}
\label{J.006}
Two jets and missing energy is one of the promising signatures of the MSSM.  In this simplified model, squarks are pair-produced and two decay topologies are explored separately. In the first decay mode, each squark decays directly to a standard model quark and the LSP, assumed to be a neutralino $\chi_1$. In the second decay topology, the squarks decay to the neutralino LSP through an intermediate chargino, $\chi^{\pm}$, producing leptons in the final state. For directly-decaying squarks, the results can be presented as  an upper limit on $\sigma_{pp\rightarrow\tilde{Q}\bar{\tilde{Q}}}\times\mathcal{B}_{\tilde{Q}\rightarrow q \chi_1}$ as a function of $(m_{\tilde{Q}}, m_{\chi_1})$. Whenever the intermediate $\chi^{\pm}$ is present in the decay chain, results can still be presented as an upper limit on $\sigma_{pp\rightarrow\tilde{Q}\bar{\tilde{Q}}}\times\mathcal{B}_{\tilde{Q}\rightarrow q' \chi^{\pm}}$ as a function of $(m_{\tilde{Q}}, m_{\chi_1})$ for different parameterizations of $m_{\chi^{\pm}}$. \\
\authorcite{J.006}{S. El Hedri, E. Izaguirre}

\subsection{Three Jets + $\MET$ from Squark-Gluino Production}
\label{J.007}
Three or more jets and missing energy is one of the signatures of the MSSM with potential for discovery in the early LHC running. In this simplified model, a squark, $\tilde{Q}$, and a gluino, $\tilde{G}$, are associatively produced. Each of the new colored particles decays to standard model states and the LSP, assumed to be a neutralino $\chi_1$. Leptons can appear in the final state whenever the squark or gluino decay to the neutralino LSP via an intermediate chargino, $\chi^{\pm}$. The exact number of jets in the final state depends on the ordering of the squark and gluino masses. Results are presented as an upper limit on $\sigma_{pp\rightarrow\tilde{Q}\bar{\tilde{Q}}}\times\mathcal{B}_i\times\mathcal{B}_j$ as a function of $(m_{\tilde{G}}, m_{\chi_1})$ for different parameterizations of $m_{\tilde{Q}}$, where $\mathcal{B}_i$ and $\mathcal{B}_j$ refer to the branching ratios of the squark and gluino, respectively. Whenever the intermediate $\chi^{\pm}$ is present in the decay chain, results can still be presented as an upper limit on $\sigma_{pp\rightarrow\tilde{Q}\bar{\tilde{Q}}}\times\mathcal{B}_i\times\mathcal{B}_j$ as a function of $(m_{\tilde{G}}, m_{\chi_1})$. Adding an extra intermediate particle results in a proliferation of free parameters in the simplified model. Simplifying parameterizations of $m_{\tilde{Q}}$ and $m_{\chi^{\pm}}$ that capture the relevant kinematics of this topology can be made. \\
\authorcite{J.007}{S. El Hedri, E. Izaguirre}

\subsection{Composite Gluon to Invisible}
\label{J.000}
This model appears in some models of strong dynamics where a composite gluon or quark decays into missing energy plus jets. The particles are the following. There is a composite gluon, $g'$, and a massive particle, $\phi$, with their respective masses. The composite gluon is produced resonantly and then decays back to the $\phi$ and a gluon. The $\phi$ is not stable but eventually decays into invisible stable particles. This model could also be seen as a dijet resonance, but the background in that channel might be sufficiently high or the resonance sufficiently broad that the decay into this channel is dominant over the dijet channel. The mass of $\phi$ is an important variable because it determines the kinematics of the $g'$ decay. The $g'$ could naturally be very broad and the decays to $g\phi$ could be very small. \\ 
\authorcite{J.000}{J. Wacker}

\subsection{Monojets from Neutralino-Squark Associated Production}
\label{J.005}
Neutralino-squark associated production is one of the lesser studied monojet channels \cite{Baer:1995nq,Baer:1990rq}.
While typically it is assumed that the quark-squark-neutralino couplings are relatively small, this does not need to be the case in more general theories. \\
\authorcite{J.005}{J. Wacker}

 \subsection{One-Stage Gluino Cascade} 
This model is summarized in Section~\ref{detailed}.

%% file: heavy.tex
\section{Heavy Flavor}\label{heavy}

Simplified models for heavy flavor are organized primarily according
to (1) whether the new physics leading to the third generation is
produced resonantly, pair-produced, or is the result of a cascade
decay, and (2) whether the new particles involved carry a flavor quantum
number or not.  The resonant models include several cases
of $Z'$ and $W'$ with enhanced couplings to the third generation (\S \ref{heavy:WZprime}, \ref{heavy:tautau}, \ref{heavy:WpN}), as well as
excited quarks (\S \ref{heavy:excited}).  Pair-production scenarios where the
produced particles do \emph{not} carry flavor quantum numbers include
gluino decays to heavy flavor (\S \ref{heavy:gluino}), pair-production of color-octet
scalars (\S \ref{heavy:scalar8}), and exotic higgs scenarios that each produce four
heavy-flavor fermions in each decay (\S \ref{heavy:4tau}).  These can produce striking final
states such as like-sign tops.  Simplified models for pair-production
of flavored ``top partner" or ``heavy $t'$ " allow several possibilities:
decays to heavy flavors + missing energy (e.g. stop, \S \ref{heavy:stop} and \S \ref{heavy:246top}), heavy flavors +
gauge bosons (vectorlike t-prime quark, \S \ref{heavy:vtprime}), or heavy-flavor quarks plus
leptons (leptoquarks, \S \ref{heavy:3glq}).  Finally, notable 3rd-generation products can
arise in decay chains through cascade decays to higgses (\S \ref{heavy:higgs}), or to
tau-partners.  Stau NLSP's in gauge mediation  (\S \ref{heavy:stau}) furnish a minimal
example of the latter possibility.
Links to the available supporting information at www.lhcnewphysics.org are given by the post number (e.g., B.008). 

\subsection{Heavy Flavor from $W'$ and $Z'$ Resonances}\label{heavy:WZprime}
Neutral vector bosons and vector bosons with charge $\pm1$ may be produced at the LHC via couplings to the light quarks and discovered through their decays to third generation fermions $t$, $b$, and  $\tau$.  A simplified model for this depends only on the resonance's couplings to fermions, its mass, and its width.  A variety of $Z'$ and $W'$ models from the literature may be mapped
onto this schema \cite{HLS,Li:1981nk,Chivukula:1994mn,Muller:1996dj,Chivukula:1995gu,Malkawi:1996fs,He:1999vp,Hill:1994hp,Georgi:1989ic,Georgi:1989xz,Carena:2004xs}.
Appropriate experimental measurements and reporting methods for the early LHC running period are discussed.  A \texttt{PYTHIA} implementation of a sequential $Z'$ boson already exists and can be used to obtain information about other $Z'$ models \cite{Ciobanu:2005pv}. \\
\authorcite{B.008, B.009}{L. Fitzpatrick, P. Ko, K. Rehermann, R. Sekhar Chivukula, M. Schmaltz, M. Schwartz,  E. Simmons, C. Spethmann,  T. Tait, W. Waltenberger}

\subsection{Tau-Tau Resonance}\label{heavy:tautau}
The relevant SU(3)$^J_Q$ quantum numbers of a $\tau^+\tau^-$ resonance are {\bf 1}$^{0,1,2}_0$.  This simplified model considers a spin-0 (pseudo)scalar $\phi$, motivated by the type-II 2HDM and/or the MSSM.  The dominant production channel is either through gluon fusion or 
$b\bar{b}\rightarrow \phi$
\cite{Dicus:1988cx,Maltoni:2003pn,Harlander:2003ai,Harlander:2002vv}.  Current limits from the Tevatron are translated into bounds on $\sigma[pp\to \phi] \times {\rm BR}[\phi \to \tau^+\tau^-]$ at the 7~TeV LHC for each of the production channels.\\
\authorcite{T.000}{A. Freitas and S. Su}

\subsection{Heavy Flavor and a Massive Right-Handed Neutrino from a $W'$}\label{heavy:WpN}
This model includes a charged scalar or gauge boson (for example, from $SU(2)_L \times SU(2)_R$ extensions of the SM electroweak interaction \cite{Mohapatra:1974hk,Mohapatra:1974gc,Senjanovic:1975rk}) resonance that is produced at the LHC through its couplings to light quarks, with subsequent decay to a lepton/tau and a heavy neutrino. The heavy neutrino then undergoes a three-body decay to a charged lepton plus two quark final state; both the lepton as well as the quarks may be from the third generation.  Monte Carlo simulation of events with this topology are discussed, and suggestions are given for how search results can be most effectively communicated. \\
\authorcite{T.003}{L. Fitzpatrick, P. Ko, K. Rehermann, R. Sekhar Chivukula, M. Schmaltz, M. Schwartz,  E. Simmons, C. Spethmann,  T. Tait, W. Waltenberger}

\subsection{Single Tops and Bottoms from an Excited Quark}\label{heavy:excited}
This simplified model consists of a single ``excited'' quark that decays predominantly to third generation quarks and is motivated by models in which the third generation experiences strong dynamics at the TeV scale
\cite{Georgi:1994ha,Suzuki:1991kh,Lillie:2007hd,Kumar:2009vs}. Excited quarks are produced in gluon-quark fusion processes in the s-channel and can have large cross sections
\cite{Cabibbo:1983bk,DeRujula:1983ak,Kuhn:1984rj,Baur:1989kv,Han:2010rf}. They decay to single top or bottom quarks and Standard Model gauge or Higgs bosons.  The model may be simulated with \texttt{PYTHIA8} as an excited B quark.\\
\authorcite{B.010}{L. Fitzpatrick, P. Ko, K. Rehermann, R. Sekhar Chivukula, M. Schmaltz, M. Schwartz,  E. Simmons, C. Spethmann,  T. Tait, W. Waltenberger}

\subsection{Multi-$t$/$b$ + $\MET$ from Gluino Decays to Heavy Flavor}\label{heavy:gluino}
This simplified model contains three new particles: a color-octet gluino, neutralino $\tilde \chi^0$, and chargino $\tilde \chi^\pm$ .  The gluino can decay to (A) $b \bar b + \tilde \chi^0$, (B) $t \bar t + \tilde \chi^0$,  or (C) $t \bar b + \tilde \chi^-$ and $b \bar t +\tilde \chi^+$ with independent branching ratios.  The decay $\tilde \chi^\pm \rightarrow \tilde \chi^0 + W^\pm$ is assumed, where the $W$ is off-shell when required by kinematics.  Such decays dominate in supersymmetric models with a higgsino LSP or heavy squarks of the first two generations
\cite{Acharya:2009gb, Baer:2009ff,Kane:2011zd}.  
This simplified model also applies to non-supersymmetric scenarios with the same decay topologies but different spins.  These topologies also roughly approximate the kinematics of gluino decays through on-shell top and/or bottom partners.  

A simple subspace of interest is when the chargino and neutralino are nearly degenerate, so that the decay products of $\tilde \chi^\pm$ are squeezed out.  This is expected whenever $\tilde \chi^0$ and $\tilde \chi^\pm$ are part of an approximate $SU(2)$ multiplet.  In this case the kinematics of each decay topology is simply parametrized by the gluino and neutralino masses.
The primary topologies of interest for searches are the symmetric ones: (A+A), (B+B), (C+C).  All topologies can give rise to a high multiplicity of $b$-tagged jets, while pairs of (B) and (C) decays can give rise to same-sign leptons and/or high lepton multiplicity.   The online note referenced below also considers cascade decays involving visible $W^\pm$ and $Z^0$ bosons. 
\\
\authorcite{B.000}{R. Essig, J. Kaplan}

\subsection{Multi-t/b Events from Pair-Produced Color Octet Scalars}\label{heavy:scalar8}
TeV-scale color octet scalars would naturally couple most strongly to the bottom and/or top quarks.     In this case, their pair production (through QCD interactions) dominates, and their decays will produce events with a high multiplicity of b-quarks in the final state: $(t\bar t) (t\bar t)$ and possibly $(b\bar b) (b\bar b)$ and $(t\bar t) (b\bar b)$ from a pair of electrically neutral octets $\Phi^0$, and $(t\bar b) (\bar t b)$ for a charged octet $\Phi^\pm$.  Indeed, a weak doublet of adjoint scalars (the $(8,2)_{1/2}$ representation of $SU(3) \times SU(2)_W$) is the only colored scalar that can have renormalizable,  minimal-flavor-violating couplings to Standard Model matter \cite{Manohar:2006ga}.  The phenomenology of these doublet models has been studied in \cite{Gerbush:2007fe}.  These considerations, and the distinctiveness of a four-heavy-quark signal, motivate a search parametrized by a very simple model.   $\Phi^0$ and $\Phi^\pm$ pair production can be considered separately, with independent masses $m(\Phi^0)$ and $m(\Phi^\pm)$.  Assuming 100\% branching fraction $\Phi^+ \rightarrow t \bar b$ ( $\Phi^- \rightarrow \bar t b$), the $\Phi^\pm$ simplified model is fully specified by $m(\Phi^\pm)$.  For $\Phi^0$ pairs, in addition to $m(\Phi_0)$, the parameter $BR(\Phi^0 \rightarrow t\bar t) = 1 - BR(\Phi^0 \rightarrow b\bar b)$ can also be varied, or simply set to 1 or 0 to study the four-$t$ or four-$b$ topologies.
Variations exist with color octet vector particles as well
\cite{Chivukula:1996yr,Simmons:1996fz,Popovic:1998vb,Hill:1993hs,Hill:1991at,Agashe:2006hk,Agashe:2003zs,Frampton:1987dn,Bagger:1987fz,Frampton:2009rk,Bai:2010dj,Chivukula:2010fk}.  \\
\authorcite{B.007}{T. Gregoire, A. Katz, N. Toro}

\subsection{Four Taus from Higgs Decays}\label{heavy:4tau}
This is a simplified model that allows for a four-tau signal, from $gg\to X \to YY\to (\tau^+\tau^-)(\tau^+\tau^-)$, with no additional missing energy except from the tau decays, and no additional jets. There is also the possibility of two taus (or perhaps $b\bar b$) and two muons, where the muons form a resonance; this latter case may actually be easier to discover and should be carefully considered. This model has motivation from the NMSSM in some limited mass regimes
\cite{Dobrescu:2000yn, Kilian:2005xt, Dermisek:2005ar, Graham:2006tr, Chang:2008cw, Lisanti:2009uy}, and from leptophilic Higgs models
\cite{Georgi:1989ic,Georgi:1989xz}.  More generally, if the Higgs sector is complicated, this signature is possible and could in some fortunate cases be of the order of picobarns. Other models, where the heavier resonance is not a scalar and is perhaps produced in $q\bar q$ collisions, or where the two tau pairs come from different particles, might also appear in this search. \\
\authorcite{T.002}{M. Strassler}

\subsection{Di-$t$/$b$ + $\MET$ from Heavy-Flavored Squarks}\label{heavy:stop}
This simplified model consists of squarks $\tilde t$ and $\tilde b$ with the charges of $t$ and $b$ quarks, a chargino $\tilde \chi^\pm$, and two neutralinos $\tilde \chi^0_1$ and $\tilde \chi^0_2$.   We consider three decay modes for $\tilde t$, with analogous decays for $\tilde b$:
\begin{enumerate}
\item $\tilde t \rightarrow t \tilde \chi^0_1 \rightarrow t + \MET $
\item $\tilde t \rightarrow b \tilde \chi^+ \rightarrow b + W + \MET$
\item $\tilde t \rightarrow t \tilde \chi^0_2 \rightarrow t + Z + \MET$
\end{enumerate}
Decays among the chargino and neutralinos are assumed to produce $W$ and $Z$ bosons (taken off-shell when required by kinematics).

An important special case of this simplified model is when decays to the second neutralino $\tilde \chi^0_2$ are ignored, and the chargino $\tilde \chi^\pm$  is nearly degenerate with $\tilde \chi^0$ so that its decay products can be neglected.  In this case, the decays 
$\tilde t \rightarrow t \tilde \chi^0_1 \rightarrow t + \MET $ and 
$\tilde b \rightarrow t \tilde \chi^- \rightarrow t + \MET $ and a simple parameter-space in terms of a stop mass, neutralino mass, and two decay modes can be considered.  
Relaxing these assumptions gives rise to noteworthy signatures such as top and bottom pairs with additional leptons from $W$ and $Z$ decays.
\\
\authorcite{B.011}{P. Fox, J. Kaplan, E. Kuflik, R. Lu, S. Mrenna, N. Toro}

\subsection{Di- Quad- and Hexa-Tops plus $\MET$ from Supersymmetry}\label{heavy:246top}
Stop squarks can play a special role in phenomenological signatures of supersymmetry.  The large top quark
Yukawa coupling leads, in many regions of parameter space, to stops substantially lighter than the other squarks.
This can enhance superpartner cascade decay branching ratios to top quarks significantly.  A simple topology that captures
this feature is for a gluino, stop, and neutralino lightest superpartner (LSP).  Over all of this parameter space where
kinematically allowed, either two, four, or six top quarks are emitted in cascade decays starting from gluino or stop pair
production, $pp \to tt,tttt,tttttt + \MET$
\cite{Acharya:2009gb}.   Because of the large top quark hadronic branching fraction, these signatures
are well suited to jets plus missing energy searches. However, the large top quark multiplicity also makes
same-sign di-lepton, tri-lepton, and four or more-lepton searches
sensitive to these signatures.  A reach or upper limit on $\sigma\times \mbox{BR}$ for
$p p   \rightarrow  {\rm jets} +  \MET$ as well as the various $p p   \rightarrow  \text{multi-lepton} +  \MET$
channels, as a function of the stop and gluino masses at fixed neutralino mass, provides a
concise summary of the sensitivity to this topology.\\
\authorcite{N/A}{R. Gray, D. Hits, S. Padhi, M. Park, S. Schnetzer, S. Somalwar, S. Thomas, Y. Zhao}

\subsection{Vectorlike Heavy Top Quarks}\label{heavy:vtprime}

Fourth-generation quarks are highly constrained within the Standard Model (SM)
\cite{Kribs:2007nz,He:2001tp,Erler:2010sk,Chanowitz:2009mz,Eberhardt:2010bm}. 
However, it is also possible that a heavy quark has vectorlike couplings to the weak interactions.  
In that case, the mass can be arbitrarily high (see \cite{DelNobile:2009st,AguilarSaavedra:2009es} for surveys of possibilities). 
Vectorlike heavy quarks are found in top seesaw models \cite{Dobrescu:1997nm,Chivukula:1998wd,He:2001fz},
Little Higgs theories \cite{Han:2005ru,Han:2003wu}, theories with
extra dimensions \cite{Agashe:2004rs,Contino:2006nn},
as well as some schemes for grand unification \cite{Bagger:1984gn,Choudhury:2001hs,Morrissey:2003sc,Kumar:2010vx}.
The decay pattern is also different from that of a fourth generation Standard Model quark.  Three simple models that illustrate this are presented:  I. a SM singlet top quark, decaying to $W$+$b$, $Z$+$t$, Higgs+$t$ ; II. a SM singlet top quark decaying to $g$+$t$, $\gamma$+$t$, $Z$+$t$ through magnetic moment couplings; III. a vectorlike heavy quark doublet including a quark of charge $+5/3$ decaying to $W$+$t$ \cite{Agashe:2006at,Contino:2008hi}. 
For these three examples, we identify the relevant parameters for a model independent analysis. \\
\authorcite{B.002}{R. Contino, R. Franceschini, M. Peskin}

\subsection{Third Generation Composite Leptoquarks and Diquarks}\label{heavy:3glq}
Composite leptoquarks or diquarks that couple predominantly to third generation quarks and leptons provide a generic \cite{Gripaios:2009dq} and spectacular signature of strongly-coupled models of electroweak symmetry breaking based on the paradigm of partial compositeness \cite{Kaplan:1991dc}, which evades problems with flavour physics constraints.  The constraints on such leptoquarks were analysed in \cite{Gripaios:2009dq}. The strongest constraints come from searches for $\mu \rightarrow e \gamma$ and $\tau \rightarrow \mu \gamma$; leptoquarks with masses as low as a couple of hundred GeV can be safe provided they couple only to quarks of a single chirality (if not, one-loop contributions to the above processes can be enhanced by putting the required helicity flip on an internal top or bottom quark). The leptoquarks couple dominantly to either $t$ or $b$ and either $\tau$ or $\nu_\tau$, and are dominantly pair-produced by QCD interactions, such that the relevant collider final states are pairwise combinations of the above. LHC search strategies and prospects for the various channels are suggested in \cite{Gripaios:2010hv}. Searches at the Tevatron in the $2b2\tau$ \cite{Abazov:2008jp} or $2b + \MET$ \cite{Abazov:2007bsa} yield bounds of 210 GeV and 229 GeV, respectively; for leptoquarks that would instead decay predominantly to top quarks the limit is around $m_t$.  \texttt{Herwig++}2.5 \cite{Gieseke:2011na} provides a full Monte Carlo implementation of composite leptoquarks. For further details, see \cite{Gripaios:2010hv}. \\
\authorcite{B.001}{B. Gripaios}

\subsection{Heavy Flavor Production from Higgs Bosons in Cascades}\label{heavy:higgs}
In this simplified model, new colored states are pair produced and decay to jets and missing energy, and, some fraction of the time, to a light (Higgs) scalar. The scalar decays predominantly to $b \bar b$ because its couplings to fermion pairs are proportional to fermion mass (a heavier scalar can also decay to top quarks and $W^{\pm}$ and $Z^0$ bosons).  This simplified model is motivated by a variety of theories, including supersymmetry
\cite{Baer:1992ef,Baer:1992kd,Kribs:2010hp}, and this topology provides an opportunity for Higgs boson discovery in early LHC searches.  Its signatures are the final states $b\bar{b} + 2n j + E_T$ and $2 b\bar{b} + 2n j + E_T$, where $n$ is an integer.  The  $b \bar b$ pairs reconstruct a resonance and the associated jets may also be heavy flavor jets. \\ 
\authorcite{B.006}{M. Buckley, P. Fox, J. Kaplan, E. Kuflik, R. Lu, S. Mrenna}

\subsection{Tau-rich Events from a Stau NLSP}\label{heavy:stau}
Low scale gauge-mediated supersymmetry breaking with significant
left-right sparticle mixing arising at large values of $\tan \beta$ can naturally
give rise to a stau slepton as the next to lightest superpartner (NLSP)
\cite{Dimopoulos:1996yq}.
Approximate flavor universality of gauge-mediation ensures that the
a selectron and smuon are slightly heavier, and decay through both charge-preserving and charge-changing reactions to the stau through the emission
of soft lepton pairs.   With this type of superpartner spectra, all cascade
decays pass through the metastable stau slepton, which decays
to the un-observed Goldstino and tau
\cite{Dutta:1998tt,Muller:1998hw,Feng:1997zr,Martin:1998vb}.  If the stau NLSP is mostly
right-handed, then the only unsuppressed cascades come from the bino
component of heaver neutralinos.  Pair production of any superpartners
with cascade decays that pass through these neutralinos then give
rise to the inclusive signature of four hard taus or two hard taus and two hard leptons,
all with missing energy, $pp \to \tau \tau \tau \tau, \tau \tau \ell \ell + \MET$.
It is important to note that since all the relevant cascade decays have the possibility
to flip the superpartner charges, these signatures arise in all charge
and lepton flavor combinations.  This signature is best covered by a di-lepton plus 
one or two identified taus plus $\MET$ search in all flavor and charge channels.
The principal strong production channels that are relevant for early LHC are pairs
of gluinos and/or squarks.  Direct weak production of charginos, neutralinos, and
sleptons will become relevant in future searches.
A reach or upper limit on $\sigma\times \mbox{BR}$ for
$p p   \rightarrow  \tau \tau \ell \ell +  \MET$ as a function of the
gluino and the chargino provides a unified summary of the
sensitivity to this topology for both strong and weak production of superpartners.
Sensitivity to the remaining soft leptons emitted in the cascades between
the selectron or smuon and NLSP stau may be illustrated in the above
parameter plane for different fixed values of the mass splitting between
these states.\\
\authorcite{N/A}{R. Gray, M. Park, S. Somalwar, S. Thomas, Y. Zhao}

%% file: leptons.tex
\section{Leptons}\label{leptons}

Lepton simplified models described in this section populate final states with up to 4+ leptons with or without new sources of missing energy. 
Populating one or two lepton channels is fairly straightforward; for example, any new $W'$ with a branching ratio to leptons gives one lepton.  Any new resonance decaying to $W$ bosons also gives one lepton. Opposite sign dileptons (2OSL) are similarly straightforward.  There are several ways to get 2SSL: for example, Majorana masses such as in MSSM gauginos or heavy neutrinos, or lepton violating couplings.  Multi-lepton signatures arise most commonly in cascade decay scenarios.  
Listed below are thirteen simplified models (not all independent) that span a wide range of topologies: 
SUSY type with neutralino or squark decays giving W or Z bosons (\S \ref{leptons:H}, \ref{leptons:D}, \ref{leptons:I}, and \ref{leptons:K}),  
same-sign leptons from sneutrino LSP (\S \ref{leptons:B}), 
multi-leptons from slepton co-NLSP with a gravitino LSP (\S \ref{leptons:A}),
multi-leptons from R-parity violation (\S \ref{leptons:C}),
tri-leptons from scalar diquarks (\S \ref{leptons:F}),
doubly-charged resonances ($H^{++}$) (\S \ref{leptons:G}), 
flavor violating scalar decays (\S \ref{leptons:E}), 
leptoquarks (\S \ref{leptons:L}), 
and leptons from spin-0 resonances that decay to $W$'s, $Z$'s, or tops (\S \ref{leptons:M}).
Links to the available supporting information at www.lhcnewphysics.org are given by the post number (e.g., L.025). 

\subsection{Multileptons from Slepton co-NLSP}\label{leptons:A}

Low scale gauge-mediated supersymmetry breaking naturally gives rise to superpartner spectra with nearly degenerate right-handed sleptons playing the role of co-next to lightest superpartner (co-NLSP), with a bino-like neutralino as the next to next to lightest superpartner (NNLSP)
\cite{Meade:2008wd,Carpenter:2008wi,Carpenter:2008he,Buican:2008ws,Rajaraman:2009ga,Ruderman:2010kj}. For spectra of this type, cascade decays from heavier superpartners always pass sequentially through the bino, then to one of the co-NLSP sleptons emitting a lepton, and finally to the un-observed Goldstino, emitting another lepton.  Therefore, pair-production of heavier superpartners gives rise to inclusive signatures that include four hard leptons and missing transverse
energy, $pp \to \ell^+ \ell^- \ell^{'+} \ell^{'-} + \MET$.   This signature is best covered by an exclusive hierarchical search for quad-leptons, tri-leptons, and same-sign dileptons, including $\MET$ in the latter two cases as
necessitated by backgrounds.  The principal strong production channels that are relevant for early LHC searches are pairs of gluinos and/or squarks.  Weak production of chargino, neutralinos, and direct production of sleptons will become relevant in future searches.
A reach or upper limit on $\sigma\times \mbox{BR}$ for $p p   \rightarrow  {\rm multi-leptons} + \MET$ as a function of the
gluino and the chargino masses provides a unified summary of the sensitivity to this topology for both strong and weak production of superpartners. \\
\authorcite{L.025}{R. Gray, M. Park, J. Ruderman, D. Shih, S. Somalwar, S. Thomas, Y. Zhao}

\subsection{Same-Sign Dileptons from Sneutrino LSP}\label{leptons:B}
Superpartner spectra with a sneutrino as the lightest superpartner 
\cite{Hebbeker:1999pi}
can give rise to interesting signatures
with leptons coming from cascade decays to the sneutrino, and missing 
transverse energy carried by un-observed sneutrinos. Inspired by these characteristics
and recent developments in the mixed sneutrino sector \cite{Thomas:2007bu}, sneutrino LSP with slepton
NLSP provides a wide range of event kinematics.  An interesting example within 
this class is obtained from spectra with squarks as the lightest strongly interacting 
superpartners above the sneutrino.
Decays of the squarks through either on- or off-shell
chargino components of the wino, or neutralino components of the wino or bino, to either the sneutrino or its
selectron $SU(2)_L$ doublet partner, give rise to leptons.  As a result, strong pair-production of gluino and/or squark
superpartners gives rise to signatures that include two leptons and missing transverse energy,
$pp \to \ell \ell + \MET$.  For cascade decays that pass through the neutralino or gluino, a large fraction of the
dilepton events are same-sign, with significantly reduced background compared with opposite-sign leptons.
A reach or upper limit on $\sigma\times \mbox{BR}$ for $p p   \rightarrow  \ell^{\pm} \ell^{'\pm} +  \MET$ as a function of the
squark or gluino and the sneutrino masses (or, alternatively, the squark or gluino and the chargino masses with fixed
sneutrino mass) provides a concise summary of the sensitivity to this topology. \\
\authorcite{N/A}{S. Arora, S. Padhi, M. Park, S. Thomas, Y. Zhao}

\subsection{$4l + \MET$ or $6l$ final states from RPV}\label{leptons:C}

Final states of four or more leptons are an interesting signature of new physics given the very low or non-existent Standard Model backgrounds. This simplified model is inspired by supersymmetry with a leptonic R-parity violating operator 
\cite{Barger:1989rk,Allanach:1999ic,Dreiner:1997uz,Dreiner:2006gu}
and accommodates topologies in which the Higgs is resonantly produced and decays to two (neutral or charged) electroweak gauginos, each of which subsequently decays to three leptons through the $LLE^c$ operator.  The decays lead to spectacular signatures with 4 leptons plus MET in the final state, or 6 leptons and no missing energy, with all flavor possibilities, including taus. \\
\authorcite{L.000}{D. S. M. Alves, J. Wacker}

\subsection{Multileptons or Same-Sign Dileptons from New Colored Particles}\label{leptons:D}

This simplified model consists of a fermionic quark and a bosonic gluon partner.  The quark partner decays to a jet and a $W^{\pm}$ or $Z^{0}$ boson, while the  gluon partner decays to a jet and a quark partner.  The new colored states are dominantly produced either in pairs or associatively.  The model leads to final states with up to four leptons, including states with two same-sign leptons, and others with no missing energy \cite{Martin:2009gi}.  Experimental searches for such quark and gluon partners are motived by Randall-Sundrum 
\cite{Randall:1999ee,Randall:1999vf,Davoudiasl:1999tf,Pomarol:1999ad,Agashe:2003zs,Agashe:2002pr} and Technicolor 
\cite{Eichten:1979ah,Dimopoulos:1979es,Dimopoulos:1979sp,Dimopoulos:1980yf,Dimopoulos:1980fj,Eichten:1986eq,Lane:1989ej,Farhi:1980xs} theories.     \\
 \authorcite{L.003}{M. Lisanti, V. Sanz}

\subsection{Same-Sign Dileptons from Maximal (Quark) Flavor Violation}\label{leptons:E}

This simplified model introduces a maximal quark flavor violating neutral scalar that only couples to the first and third
generations ~\cite{BarShalom:2007pw}.  The signature of this model is an excess of same-sign top quarks events~\cite{BarShalom:2008fq}, which is most readily observed in same-sign dilepton events produced when both tops decay leptonically.
Existing constraints in the $\ell^{\pm} \ell^{\pm} b \MET$ event sample from CDF are presented in~\cite{Aaltonen:2008hx}. \\
\authorcite{L.006}{F. Yu}

\subsection{Same-Sign Trilepton Signature from Scalar Diquarks}\label{leptons:F}

A simplified model is presented that includes two scalar diquarks, which may transform as either triplets or sextets under $SU(3)$ color, and a heavy $Z'$ with flavor-off-diagonal couplings to right-handed up-type quarks \cite{Hewett:1988xc,Chen:2008hh,Shu:2009xf,Mohapatra:2007af}.  This model can yield a striking signature including three high-$p_T$ charged leptons, all with the same sign, plus jets and missing energy.     \\
\authorcite{L.014}{T. Okui, B. Thomas}

\subsection{Same-Sign Dileptons from a Resonance}\label{leptons:G}

A simplified model for a same-sign dilepton resonance is presented.  The relevant ${\rm SU(3)}_Q^J$ quantum numbers are ${\bf 1}_2^{0,1,2}$. For simplicity, only a spin 0 scalar, which is typically referred to as a doubly charged Higgs $H^{\pm\pm}$ in the literature, is considered
\cite{Gunion:1996pq}.   
The production channels, current constraints from direct and indirect searches, and LHC discovery potential are considered for the three simplest cases where $H^{\pm\pm}$ resides in a singlet, doublet or triplet ${\rm SU(2)}_L$ representation. 
If $H^{\pm\pm}$ resides in an ${\rm SU(2)}_L$ representation with a neutral component that obtains a non-zero vacuum expectation value
$\langle H^0 \rangle$, a $H^{\pm\pm}-W^{\mp\mp}-W^{\mp\mp}$ coupling arises, which is proportional to $\langle H^0 \rangle$.  Such a doubly
charged Higgs can also be considered as a same sign $W^\pm W^\pm$ resonance.  \\
\authorcite{L.018}{S. Su, V. Rentala}

\subsection{Leptons from Gluino Cascades}\label{leptons:H}

Particles with gluon quantum numbers are motivated by a variety of beyond the Standard Model scenarios, among which low-energy supersymmetry is the most popular.  This simplified model consists of four new particles, which have the quantum numbers of a gluino, a chargino, and the lightest two neutralinos.  The new states carry parity and the lightest neutralino is stable.  The gluon partners are pair produced and cascade decay through either the neutralinos or charginos, resulting in jets, missing energy, and a $W^{\pm}$ or $Z^0$ boson, which decays leptonically
\cite{Baer:1986au,Alwall:2008va,Izaguirre:2010nj,Alves:2010za}.  Limits can be parameterized in terms of the masses of the new particles and the production cross section for the gluon partners. \\
\authorcite{L.008}{E. Izaguirre, P. Schuster, N. Toro, J. Wacker}

\subsection{Multileptons from Electroweakino Production}\label{leptons:I}

Multileptons can result from pair-production of electroweakly charged objects that frequently occur in BSM theories as partners of electroweak and Higgs bosons (i.e., MSSM, NMSSM, UED, and some Little Higgs models).  This simplified model consists of  an electrically charged particle accompanied by two neutral particles, the lighter of which is stable.  In particular, the particles have the quantum numbers of the lightest two neutralinos and the lightest chargino.  The new particles are produced via Standard Model vector bosons.  Several of the production modes and decay spectra result in multilepton topologies
\cite{Baer:1986vf,Baer:1992dc,Baer:1993tr,Baer:1994nr,Matchev:1999yn,Matchev:1999nb}.  \\
\authorcite{L.029}{J. Kaplan, M. Lisanti, P. Schuster, T. Tait, J. Wacker}

\subsection{Same-Sign Dilepton + MET + X Minimal Model}\label{leptons:J}

This simplified model is constructed to analyze the same-sign dileptons (2SSL) + MET + X signature at the LHC. 
It is ``minimal'' in the sense that no new particles are introduced other than those needed to produce this signature via strong interactions with missing energy given by a new stable particle with $L = 1$.  The model contains four new particles: a gluino $\tilde{g}$, a squark $\tilde{q}$, a ``pseudo-chargino" $\tilde{D}$, and a sneutrino $\tilde{\nu}$. 
The model respects all the Standard Model gauge and global symmetries, as well as an R-parity $Z(2)_R$ under which the new particles are odd.  The predictions can be completely described in terms of three parameters: the masses of the $\tilde{g}$, $\tilde{D}$, and $\tilde{\nu}$.  For generic parameters, the CDF inclusive search 
\cite{Abulencia:2007rd} for 2SSL+MET (with 1 fb$^{-1}$ integrated luminosity) rules out gluino masses below 240 GeV at 95\% confidence.  The LHC should  substantially improve this reach with the 2010-12 run due to the large (gluino-pair) production cross section and the clear 2SSL+MET signature that appears with a 100\% branching ratio.  If a signal is observed, the collider variable $M_{CT2}$ can be used to determine the masses of the new particles. \\
\authorcite{L.001}{J. Berger, W-S. Cho, M. M. Nojiri, M. Perelstein}

\subsection{Same-sign Dileptons with Neutralino-like LSP}\label{leptons:K}

This simplified model produces a same-sign dilepton signature. Two versions of the model are proposed.  The minimal version consists of a quark and gluon partner and is based on Randall-Sundrum or Technicolor theories.   The second version supplements these new colored states with a chargino-like and neutralino-like particle; the new particle parity imposed in this case is based on scenarios such as Supersymmetry and Universal Extra Dimensions.  Both variants produce missing energy, but in the former, the missing energy solely arises from a leptonic $W^{\pm}$, whereas there is an additional contribution from the new stable particle in the latter
\cite{Martin:2009gi,DeSimone:2009ws}. \\
\authorcite{L.020}{V. Sanz}

\subsection{Leptoquarks}\label{leptons:L}

Leptoquarks are boson fields mediating lepton-quark interactions \cite{Buchmuller:1986zs}
and are predicted in a wide range of theories beyond the Standard Model that place leptons and quarks on an equal footing, such as Grand Unified theories, R-parity violating SUSY models, and composite models of leptons and quarks. Leptoquarks have a high discovery potential at the early stage of the LHC, because they can be produced via the strong QCD interactions and also give rise to clean leptons in the final state \cite{Hewett:1993ks,Hewett:1989cs,Hewett:1987yg,Hewett:1987dq,Blumlein:1996qp,Kramer:1997hh,Kramer:2004df}. \\
\authorcite{E.001}{Y. Bai, H-C. Cheng, J. Hewett}

\subsection{Multiple Weak Bosons from Strong Spin-0 Resonances}\label{leptons:M}

This simplified model captures signals arising from the top quark
coupled to a strong electroweak symmetry breaking sector. These
are relevant for technicolor, topcolor, composite Higgs and 5D ``holographic" versions of these models, in which the top quark mass
arises from a coupling to an operator with the quantum numbers of
the Higgs. These theories contain spin-0 resonances that decay to
Ws, Zs, tops, and (if present in the model) a light composite Higgs.
These signals are modeled by a non-supersymmetric
2 Higgs doublet model with custodial symmetry. 
Signals include $W^+W^-Z$, $Z h$, $Z t \bar{t}$ and $4 W$ \cite{Evans:2009ga}.
A 2HDM4TC Madgraph
package (with model and calculator) is provided.\\
\authorcite{N/A}{S.~Chang, J.~Evans, M.~Luty}

%% file: photons.tex
\section{Photons}\label{photons}

Simplified Models for photons are organized by 
the number of photons and leptons in the final state, and by whether
or not there is large MET (e.g. 2 photons with MET, 3 photons without
MET, etc). Preference was given to simplified models with colored particle production, 
appropriate for relatively early analyses at the LHC.  
However,  simplified models with electroweak production are also included because isolated photons are very
clean signatures.
Currently, the four classes of simplified models supported with write-ups include:
photon production from SUSY neutralino decays to gravitinos (\S \ref{photons:A}),
photon or other gauge boson production in the cascade decays of new resonances (\S \ref{photons:B}),
photons from the decay of exotic particles that do not involve missing energy (\S \ref{photons:C}),
and di-bosons (photon) production from the decay of a new resonance (\S \ref{photons:D}).
Links to the available supporting information at www.lhcnewphysics.org are given by the post number (e.g., P.000). 

\subsection{Diphotons from a Neutralino NLSP}\label{photons:A}

Two photons plus missing energy is one of many possible signatures of general gauge mediation models, arising when the lightest neutralino is bino-like and decays to its superpartner and a gravitino. Another possible signature is $\gamma+$lepton plus missing energy, which arises when the neutralino is wino-like. The principal production channels at the early LHC are gluino or squark pair production (or associated production). An upper limit on $\sigma\times \mbox{BR}$ for $\tilde g \tilde g ( \tilde q \tilde q)   \rightarrow \text{jets} + \gamma \gamma + \MET$, $\text{jets} + \ell \gamma + \MET$ as a function of the gluino (squark) and neutralino masses provides a concise summary of the sensitivity of this search \cite{Ruderman:2011vv}. \\
\authorcite{P.000}{Y. Gershtein, M. Park, J. Ruderman, D. Shih, S. Thomas, Y. Zhao}

\subsection{Multiphoton Production from an Intermediate Resonance}\label{photons:B}

This model consists of two new particles: a spin-0 and spin-1 SU(2)$_{\text{L}}$ triplet
(for examples, see i.e. \cite{Bai:2010qg,Bai:2010mn}).  These particles are either pair-produced via Drell-Yan  or produced from the decay of a single resonance formed by $q$-$\bar{q}$ fusion.  The decays lead to final states with four electroweak gauge bosons.  This model provides a simple benchmark for multiphoton production, with or without jets, with or without missing energy.   \\
\authorcite{P.002}{T. Okui}

\subsection{Photons without $\MET$ from an Exotic Octet}\label{photons:C}

This model consists of a new real scalar in the adjoint representation of the color gauge group and gives the signature of photon(s) + X.  It is an IR effective theory of a broad class of models, including supersymmetry with Dirac gauginos \cite{Polchinski:1982an,Hall:1990hq,Randall:1992cq,Dine:1992yw,Fox:2002bu,Kribs:2007ac,Plehn:2008ae}
and models with strong dynamics.  Most of the events with isolated photons do not have missing energy, a tell-tale signature that distinguishes this scenario from gauge mediation models with the gravitino as the lightest supersymmetric particle.  \\
\authorcite{P.003}{T. Okui, T. Roy}

\subsection{Dibosons from a Resonance}\label{photons:D}

This model introduces new resonance states that decay to di-bosons, including $g+ g$, $g + \gamma$, $g + W^\pm$, $\gamma + \gamma$, $\gamma + Z$, $Z + Z$ and $Z + W^\pm$.  These resonances carry an approximate $Z_2$ symmetry, which is broken by higher dimensional operators induced by anomalous Wess-Zumino Witten interactions.  Having a very narrow width, these resonances are dominantly pair-produced at the LHC from gauge interactions, and result in a rich final state of multiple standard model gauge bosons.  The signatures and bounds for the most promising channels ($\gamma+\gamma$, $\gamma+g$, and $\gamma+W^\pm$) are considered \cite{Bai:2010mn,Freitas:2010ht}. \\
\authorcite{P.004}{Y. Bai, J. Evans, A. Freitas, P. Schwaller}


%% file: exotica.tex
\section{Exotica}\label{exotica}

Exotica simplified models are organized around the production of  ``exotic objects'' that do not manifest themselves as standard jets, 
leptons, or missing transverse energy (MET).  
Such objects include, for example, charged massive particles (CHAMPS) (\S \ref{exotica:A}) and ``lepton-jets'' (\S \ref{exotica:D}), 
as well as objects that produce ``weird'' tracks (e.g.~kinks or intermittent tracks) (\S \ref{exotica:A}) or high-multiplicity tracks (\S \ref{exotica:E}).  
We also include simplified models with displaced vertex signatures that arise from new resonance production (\S \ref{exotica:B}) or
appear in association with jet production (\S \ref{exotica:C}).
Simplified models that produce signatures that could be missed at the trigger level are also discussed (\S \ref{exotica:A}).
All of these exotic objects and signatures are found in various scenarios for new physics, 
including Hidden Valley scenarios, gauge-mediated supersymmetry breaking, supersymmetry with R-parity violation, 
quirks etc.  
Links to the available supporting information at www.lhcnewphysics.org are given by the post number (e.g., E.005). 

\subsection{Unusual Energy Deposition, Timing, and Tracks} \label{exotica:A} 

A simplified model based on long lived charged particles is proposed, which results in many possible signatures beyond those considered in standard CHAMP/HSCP searches.  The signatures include reduced or increased energy deposition (``$dE/dx$'') in the HCAL, ECAL, or $\mu$-chamber, anomalous timing as measured in various detector components, or irregular tracks such as kinks and intermittent hits \cite{Meade:2011du}.
This diverse array of signatures can be modeled using only two particles and a small number of parameters.  This simplified model captures a wide variety of new physics models, such as gauge-mediated supersymmetry breaking (GMSB) \cite{Giudice:1998bp}, 
quirks \cite{Kang:2008ea}, supersymmetry with R-parity violation (RPV)
\cite{Meade:2011du,Berger:2000mp}, split-supersymmetry
\cite{ArkaniHamed:2004yi,ArkaniHamed:2004fb}, and  monopoles \cite{Ginzburg:1998vb}.  
Because these signatures are not standard, they may be missed by triggers for standard searches.  \\
\authorcite{E.005}{R. Essig, P. Meade, J. Shao, T. Volansky, I. Yavin}

\subsection{Displaced Vertices from a Resonance} \label{exotica:B} 

Displaced vertices arise in many well-motivated extensions of the Standard Model.
 While their appearance can have important implications for fundamental physics,
they also present a particular challenge to experimental identification
that has not been fully explored.  This simplified model contains a resonant connector particle
that decays to a pair of long-lived states (such as in Hidden Valley models
\cite{Han:2007ae,Strassler:2006im}).  These long-lived modes eventually decay back to the Standard Model, potentially producing one or two displaced vertices.  The decay products at these vertices are chosen from theoretically motivated scenarios, covering a wide range of allowed two and three-body modes into the Standard Model.     \\
\authorcite{E.008}{S. Chang, A. Haas, D. Morrissey}

\subsection{Displaced Vertices with Associated Jets} \label{exotica:C} 

Many well-motivated extensions of the Standard Model give rise to collider
signals consisting of displaced vertices together with hard QCD jets.
Signals of this type can have important implications for fundamental
physics, but they also present a significant challenge to experimental
identification.  This simple model describes displaced vertex signals
that can arise from long-lived states produced in association
with hard  QCD jets.  This occurs, for example, when a new colored state is produced,
which subsequently decays into a quark and a long-lived particle (such as a
long-lived neutralino, in supersymmetric models). These long-lived modes
eventually decay back to the Standard Model, potentially producing one or
two displaced vertices. The decay products at these vertices are chosen from
several theoretically motivated scenarios, covering a wide range of allowed
two and three-body modes into the Standard Model. \\
\authorcite{E.010}{S. Chang, A. Haas, D. Morrissey}

\subsection{Weird Jets} \label{exotica:D} 

This simplified model gives rise to the production of jet-like structures that differ from ordinary QCD jets.  In comparison to standard jets, these new objects differ in their particle constituents (i.e., jets composed primarily of leptons or photons
\cite{ArkaniHamed:2008qp,Cheung:2009su,Bai:2009it}), in the multiplicity of the constituents, and/or in their transverse/longitudinal shower profiles \cite{Butterworth:2008iy,Almeida:2008tp,Salam:2009jx,Abdesselam:2010pt}.  These properties can be easily controlled by two mass scales determining the mass 
and shape of the jet, and a set of branching ratios and lifetimes into Standard Model final states determining the jet composition.  A small set of simplified models is sufficient to parameterize a wide variety of experimental signatures arising in various models of
new physics such as Hidden Valleys and extended Higgs sectors with new light states. \\
\authorcite{N/A}{D. Krohn, M. Papucci, D. Phalen}

\subsection{High Multiplicity} \label{exotica:E} 

High multiplicity signatures, with the exception of scenarios in which new physics appears as a modification to the underlying event, are characterized by both high multiplicity and high $H_T$.  There are several sub-signatures that fall under this general umbrella definition:
\begin{itemize}
\item Very high multiplicity ``thermal" distributions arising from TeV  
black hole evaporation \cite{Dimopoulos:2001hw,Giddings:2001bu,Rizzo:2005fz}.
\item Spherical events such as those arising in conformal hidden  
sector physics models, as in Hidden Valleys \cite{Han:2007ae,Strassler:2006im}
or Unparticle \cite{Georgi:2007si,Georgi:2007ek} models.
\item A high multiplicity of sub-weak scale resonances decaying into  
pairs of SM particles.
\item High multiplicities of SM particles with either large or small  
amounts of transverse missing energy, such as those arising from extra-  
long SUSY cascades into a  hidden sector.
\end{itemize}
This simplified model parameterizes the production and decay of hidden sector  
particles through higher dimension operators.  To cover the many sub-signatures, the following searches, at a minimum, are suggested:
\begin{itemize}
\item di-lepton resonance plus anything in a high $H_T$ event;
\item di-photon resonance plus anything in a high $H_T$ events;
\item high $H_T$ with reduced missing energy in extended SUSY decay  
chains;
\item multi-lepton, multi-jet high $H_T$ events, where weak $p_T$ cuts  
on the jets, leptons and photons are traded for high multiplicities of  
objects.
\end{itemize}
\authorcite{E.004}{M. Baumgart, J. Hubisz, K. Zurek}

%% file: conclusions.tex
\section{Summary and Outlook}

Searching effectively for new physics in the data-rich LHC environment requires posing specific questions about interesting classes of events, rather than simply casting a wide net.  The most meaningful and effective search results (either positive or negative) are those that inform future searches as to relevant
parameter spaces, that can be usefully compared to similar searches by different experiments, and that allow a transparent interpretation in 
terms of particle production and decay. 
 
Interpreting search results in the context of few-particle simplified models can facilitate all three goals.  To this end, we have proposed a catalog of simplified
models, covering a wide variety of models and new-physics signatures.  These simplified models are not intended to displace signature-based analyses, or the interpretation of search result within other specific models (e.g. mSUGRA limits), 
but rather to complement these results with a different emphasis.

Focusing on a small relevant set of particles and interactions at a time allows the design of searches that are robust over a more complete range of masses and spectral possibilities -- for example, decay chains or kinematic regions that  may never occur in a particular model sub-space, but require a distinct search design.  The weaknesses of one search's simplified model coverage become parameter ranges to focus on for future revisions of the search. 
Simplified models have the further virtue that they allow an interpretation for signature-based searches, even when the signatures do not arise in existing, theoretically motivated models.  

An important virtue of searches with a model-based interpretation relative to pure signature-based searches is that they can be compared across collider experiments.  Simplified models provide a figure of merit for comparing searches at different collider experiments, because the kinematics and cross-sections expected for a simplified model at different colliders can be computed from their fundamental parameters.  Though it is often useful to view production cross-sections as free 
parameters, the couplings in a simplified model allow a computation of \emph{expected} cross-sections in different situations, allowing comparison of results among the Tevatron  and various LHC runs.
 
A balance must be achieved in the interpretation of LHC search results and new-physics models.  One would like for searches to maximize coverage of popular visions for physics Beyond the Standard Model, but also to be applicable to a broad range of related models.  In particular, when the application of a search to related models is in principle straightforward, it should not require re-analysis by the experiment.  Simplified models are described by parameters that are closely related to experimental signals, and less particular to an individual model that inspired a search.   Limits on simplified-model topologies are directly and simply applicable to other models that share the same topologies.  Though these limits are often approximate and in some cases significantly weaker than the limits that could be derived directly, they represent a significant improvement over typical model-specific results, which permit no reliable translation of results into other models of interest. Moreover, a characterization 
of the search results directly in terms of physical masses and topology is typically transparent and instructive.

The catalogue of models in this note (largely developed as part of the SLAC workshop) is by no means complete.  We hope it provides a baseline for analyses in early data sets, and useful suggestions for presentation of results.  But the list will necessarily evolve with time.  Various models have been omitted from this note because their expected cross-sections are quite low, but they will be quite interesting when multiple inverse femtobarns have been recorded.  Moreover, future discoveries at the LHC will undoubtedly motivate a much more refined study of particular models and parameter regions, including detailed study of correlated channels and development of complete effective theories to describe the observed dynamics.  On the road to discovery and understanding of physics beyond the Standard Model at the LHC, simplified models represent
a powerful and interesting step.

%% file: acknowledgments.tex
\acknowledgments

We acknowledge the participants of the SLAC and CERN workshops for lively and productive discussions.
We thank Joe Incandela, Cigdem Issever, Paul de Jong, Michelangelo Mangano, and Fedor Ratnikov for organizing the CERN workshops 
on ``Characterization of New Physics at the LHC'' I and II  as well as feedback, support, and valuable discussions during 
the completion of this work. We also thank Claudio Campagnari, Kyle Cramner, Albert De Roeck, Amir Farbin, Steven Giddings, Louise Heelan, Boaz Klima, Zach Marshall, Jeff Richman, Roberto Rossin, David Stuart, and Chris Tully for support and valuable discussions. 
The Workshop on ``Topologies for Early LHC Searches'' was supported by SLAC National Accelerator Laboratory's High Energy Theory Group, Department of Particle Physics and Astrophysics, and J. Wacker's Outstanding Junior Investigator Award and Sloan Fellowship Award.  
Rouven Essig is supported by the US DOE under contract number DE-AC02-76SF00515.  
Tim M.P.~Tait is supported by the NSF under grant PHY-0970171.
Mariangela Lisanti acknowledges support from the LHC-TI and Simons Fellowships.
Spencer Chang, Jared A.~Evans, and Markus Luty are supported under DOE Grant \#DE-FG02-91ER40674.
The work of R.~Sekhar Chivukula and Elizabeth H.~Simmons is supported in part by the National Science Foundation under award PHY-0854889. 
Sanjay Padhi acknowledges support from the DOE under the grant DOE -FG02-90ER40546.
Maxim Perelstein is supported by the U.S. National Science Foundation through grant PHY-0757868 and CAREER award PHY-0844667. 
The work of Can Kilic is supported by DOE grant DE-FG02-96ER50959.
Pedro Schaller acknowledges support from the DOE (HEP Division) grants DE-AC02-06CH11357
and DE-FG02-84ER40173.
Matthew D.~Schwartz acknowledges support from DOE grant DE-SC003916. 
Felix Yu was supported in part by the National Science Foundation under the
grants No. PHY-0653656 and PHY-0970173.  Felix Yu was also supported by a 2010
LHC Theory Initiative Graduate Fellowship, NSF Grant No. PHY-0705682.